\documentclass[aps,prl,twocolumn,superscriptaddress,nofootinbib]{revtex4-2} % added handling of footnotes
\usepackage{chemformula} % chemical equations
\usepackage{longtable} % tables with pagebreak
\usepackage{multirow} % combine cells in table
\usepackage{amssymb} % additional mathematic symbols
\usepackage{enumitem} % allow lists without spacing
\usepackage{float} % better placement of figures
\usepackage{pdfpages} % include PDFs into the file
\usepackage{pgffor} % iterate over pdf file
\makeatletter
\AtBeginDocument{\let\LS@rot\@undefined}
\makeatother

\begin{document}

\title{Relaxation effects in twisted bilayer molybdenum disulfide: structure, stability, and electronic properties}

\author{Florian M. Arnold}
    \affiliation{TU Dresden, Theoretical Chemistry, Bergstr. 66c, 01062 Dresden, Germany}
\author{Alireza Ghasemifard}
    \affiliation{TU Dresden, Theoretical Chemistry, Bergstr. 66c, 01062 Dresden, Germany}
\author{Agnieszka Kuc}
\email{a.kuc@hzdr.de}
    \affiliation{Helmholtz-Zentrum Dresden-Rossendorf, Abteilung Ressourcenökologie,Forschungsstelle Leipzig, Leipzig, Germany}
\author{Jens Kunstmann}
    \affiliation{TU Dresden, Theoretical Chemistry, Bergstr. 66c, 01062 Dresden, Germany}
\author{Thomas Heine}
    \email{thomas.heine@tu-dresden.de}
    \affiliation{TU Dresden, Theoretical Chemistry, Bergstr. 66c, 01062 Dresden, Germany}
    \affiliation{Helmholtz-Zentrum Dresden-Rossendorf, Abteilung Ressourcenökologie,Forschungsstelle Leipzig, Leipzig, Germany}
    \affiliation{Yonsei University and ibs-cnm, Seodaemun-gu, Seoul 120-749, Republic of Korea}

\date{\today}

\begin{abstract}
Manipulating the interlayer twist angle is a powerful tool to tailor the properties of layered two-dimensional crystals.
The twist angle has a determinant impact on these systems' atomistic structure and electronic properties.
This includes the corrugation of individual layers, formation of stacking domains and other structural elements, and electronic structure changes due to the atomic reconstruction and superlattice effects.
However, how these properties change with the twist angle, $\theta$, is not yet well understood.
Here, we monitor the change of twisted bilayer MoS\textsubscript{2} characteristics as a function of $\theta$. 
We identify distinct structural regimes, each with particular structural and electronic properties.
We employ a hierarchical approach ranging from a reactive force field through the density-functional-based tight-binding approach and density-functional theory.
To obtain a comprehensive overview, we analyzed a large number of twisted bilayers
with twist angles in the range of $\theta=0.2^\circ\dots59.6^\circ$.
Some systems include up to half a million atoms, making structure optimization and electronic property calculation challenging.
For $13^\circ \lessapprox \theta \lessapprox 47^\circ$, the structure is well-described by a moiré regime composed of two rigidly twisted monolayers.
At small twist angles ($\theta\leq3^\circ$ and $57^\circ\leq\theta$), a domain-soliton regime evolves, where the structure contains large triangular stacking domains, separated by a network of strain solitons and short-ranged high-energy nodes.
The corrugation of the layers and the emerging superlattice of solitons and stacking domains affects the electronic structure.
Emerging predominant characteristic features are Dirac cones at $K$ and kagome bands.
These features flatten for $\theta$ approaching $0^\circ$ and $60^\circ$.
Our results show at which range of $\theta$ the characteristic features of the reconstruction, namely extended stacking domains, the soliton network, and superlattice, emerge and give rise to exciting electronics.
We expect our findings also to be relevant for other twisted bilayer systems.
\end{abstract}

\maketitle

\section{Introduction}

\begin{figure*}[ht!]
    \centering
    \includegraphics[width=0.8\textwidth]{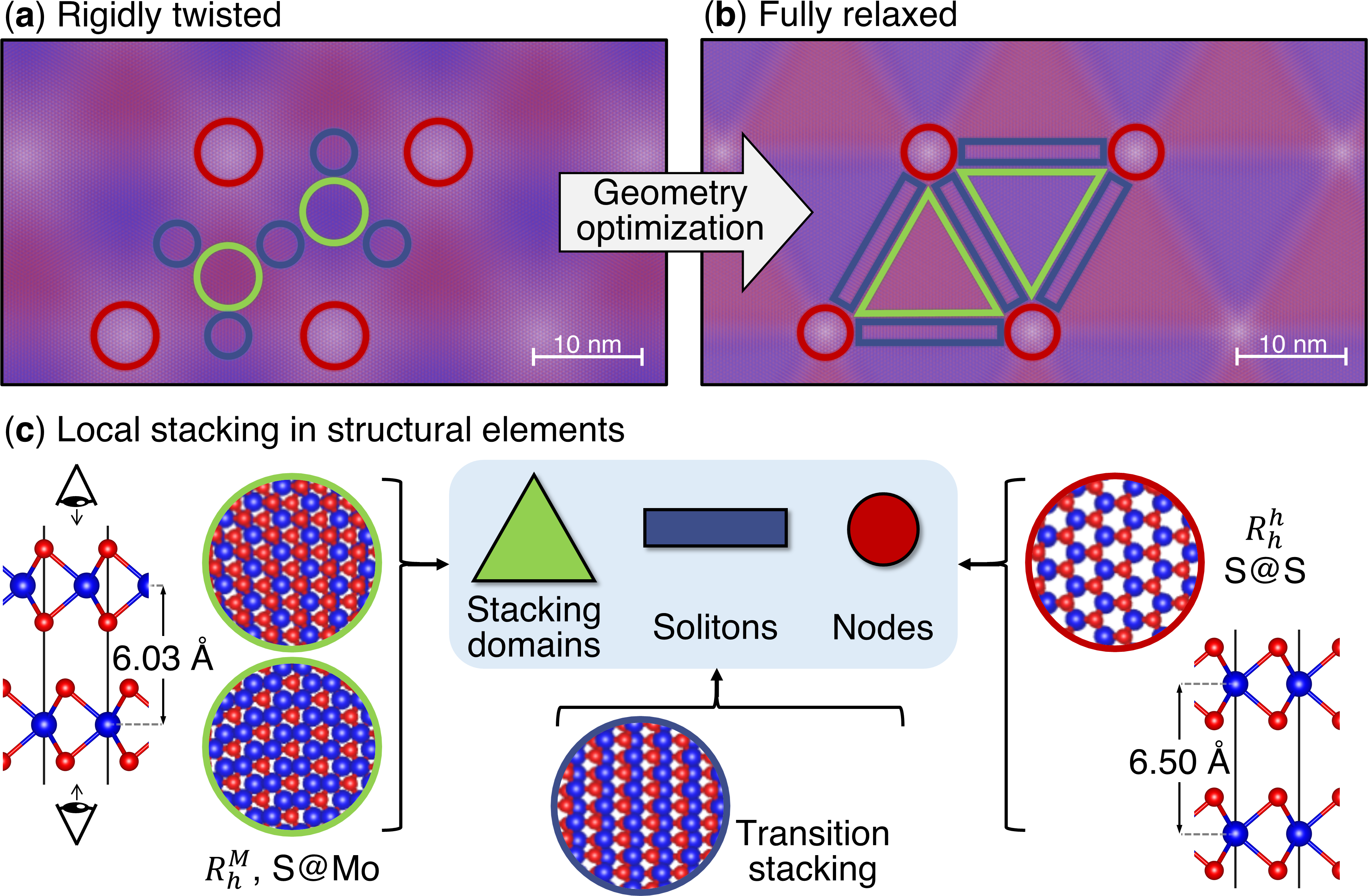}
    \caption{Atomic structure of a twisted bilayer (tBL) MoS\textsubscript{2} system with a twist angle $\theta=1.05^\circ$ in the case of (a) rigidly twisted and (b) fully relaxed layers (Mo - blue, S - red). Areas of local high-symmetry stacking are indicated. (c) Structural elements and their respective local stacking for $\theta\rightarrow0^\circ$. The stacking domains (green triangles) exhibit local $R_h^M$ stacking, either seen from the top (S-rich side) or the bottom (Mo-rich side). The solitons (blue rectangles) show a transition stacking type. The nodes (red circles) have local $R_h^h$ stacking, with eclipsed S atoms. The local stackings for $\theta\rightarrow60^\circ$ are shown in the SI (Fig.~S1).}
    \label{fig_schematic_relaxation}
\end{figure*}

The idea of combining two-dimensional (2D) crystals by stacking opened up an intense research focus in the emergent field of van der Waals (vdW) [hetero]structures.\cite{geim2013vdW-HS}
The ability to control the stacking order and to introduce an interlayer twist angle, $\theta$, gives seemingly endless possibilities to tune the properties of these materials.\cite{vanderZande2014tailoring,vanWijk2015,lu2017twistedheterobilayers,dong2021highthroughput,vitale2021tBLtmdc}
A consequence of this is the formation of moiré patterns,~\cite{constantinescu2015mos2mose2,carr2018relaxation,ribeiro2018twistable,yoo2019graphene} which exhibit a continuous change of the local stacking as visualized in Fig.~\ref{fig_schematic_relaxation}(a).
In particular for small twist angles near $0^\circ$ and $60^\circ$, domains of low-energy high-symmetry stackings are formed.
They act as seeds for a relaxation process, where energy minimization yields a maximization of the size of low-energy domains, as visualized in Fig.~\ref{fig_schematic_relaxation}(b).
Relaxation affects the in-plane (ip) structure, forming the stacking domains as dominating structural elements with local high-symmetry stacking (see Fig.~\ref{fig_schematic_relaxation}(c)).
By corrugation, also the out-of-plane (oop) structure is affected.
So-called strain solitons~\cite{kumar2016limits,gargiulo2017graphene,gadelha2021localization,lamparski2020soliton}, with the nodes at their intersections, dominate the corrugated areas.
Similar effects have been observed in simulations of a range of different twisted bilayer (tBL) materials, such as graphene~\cite{vanWijk2015,gargiulo2017graphene}, MoS\textsubscript{2}~\cite{vanderZande2014tailoring,maity2020phononsTMDCs}, or MoS\textsubscript{2}/WSe\textsubscript{2}~\cite{kunstmann2018momentum}.
Several recent works also give strong experimental evidence of such reconstruction effects.\cite{yoo2019graphene,weston2020tBLtmdcs,zhao2023hoegele}

Depending on the value of $\theta$ ($0^\circ\leq\theta\leq60^\circ$), the characteristics of the systems change, and different structural regimes are obtained:
A $\theta$ of exactly $30^\circ$ results, as in bilayer graphene, in a quasicrystalline arrangement.\cite{ahn2018quasicrystal}
For large twist angles ($\theta\rightarrow 30^\circ$), in the so-called moiré regime~\cite{yoo2019graphene}, structural relaxation of individual layers is minor.\cite{maity2020phononsTMDCs,gargiulo2017graphene}
Here, the weak oop deformation exhibits a sinusoidal modulation and decreases further for $\theta\rightarrow30^\circ$.\cite{vanWijk2015,dai2016twistedgraphene,vanderZande2014tailoring}
At small twist angles, $\theta \rightarrow 0^\circ$ and $\theta \rightarrow 60^\circ$, the structure is commonly described by a (strain) soliton regime, where an extended atomic reconstruction both ip and oop takes place.\cite{yoo2019graphene,kumar2015elastic,naik2019KCSW,shabani2021deep}
This results in a regular pattern in the interlayer distance landscape which can be understood in terms of structural elements.\cite{zhu2018moire}
The first structural element is the so-called stacking domain -- large areas of minimum interlayer distance and local low-energy stacking, marked with green triangles in Fig.~\ref{fig_schematic_relaxation}(b) for $\theta\rightarrow0^\circ$ (see Fig.~S1 for comparison to $\theta\rightarrow60^\circ$).
They are separated by domain walls of transition stacking, usually called solitons or strain solitons, as a second structural element.\cite{kumar2016limits,gargiulo2017graphene,gadelha2021localization,lamparski2020soliton}
The solitons are marked by blue rectangles in Fig.~\ref{fig_schematic_relaxation}(b) and act as a network of domain walls, separating adjacent stacking domains and vanishing only at exactly $0^\circ$ or $60^\circ$.
The stacking domains form a superlattice with honeycomb (\textit{hcb}) symmetry, while the centers of the solitons form a kagome (\textit{kgm}) lattice.\cite{angeli2021macdonald,arnold2023chemRxiv}
As the final structural element, the so-called nodes are formed at the intersections of the solitons, marked with red circles in Fig.~\ref{fig_schematic_relaxation}(b).
These are areas with maximum interlayer distance and local high-energy stacking.
The separation of the nodes correlates with the moiré periodicity.
The twist angles at which the transition between structural regimes take place depend on which system is studied, e.g., $\approx1^\circ$ in tBL graphene~\cite{gargiulo2017graphene,yoo2019graphene,dai2016twistedgraphene}, $\approx 4^\circ$ in tBL WSe\textsubscript{2}~\cite{wang2020tmdcs}, and up to $13^\circ$ in tBL MoS\textsubscript{2}~\cite{maity2020phononsTMDCs}.
However, the criteria to define these transition angles are not uniquely defined in the literature.
Different reports use deformation energy, moiré periodicity, or the size of structural elements, amongst others.

Introducing a small interlayer twist into a homobilayer system results in several hundreds or thousands of atoms in the moiré supercell.
While at $\theta=2^\circ$, there are 5000 atoms, this number increases to 20,000 at $1^\circ$ and reaches more than 100,000 for twist angles below $0.44^\circ$.
This makes simulations challenging and calls for approximate methods.
Thus, to date, the literature is restricted to results from continuum mechanics~\cite{kumar2016limits} and force field~\cite{ostadhossein2017reaxff,naik2019KCSW} approaches.

In this work, we elucidate how the structure and electronic properties (band structure, local density of states (LDOS), local gap, and electronic transport) of tBL MoS\textsubscript{2} change with the twist angle, $\theta$.
To provide a complete picture, we obtain our results for a very large set of systems, allowing us to describe structural changes with $\theta$ almost continuously.
We derive a complete characterization of the structural regimes in tBL MoS\textsubscript{2} and the transition between these regimes.
We analyze the strong atomic reconstruction for very small $\theta$ and the emergence of Dirac cones and kagome bands in the electronic structure associated with developing superlattices of the structural elements.
These features can only be visible for fully relaxed structures and are missing in band structures of rigidly twisted layers.
This calls for atomic relaxation, especially for very small $\theta$.
Our approach can be adapted to other twisted vdW structures where atomic relaxation effects play a role.

\section{Methods}

\begin{figure*}[ht!]
    \centering
    \includegraphics[width=\textwidth]{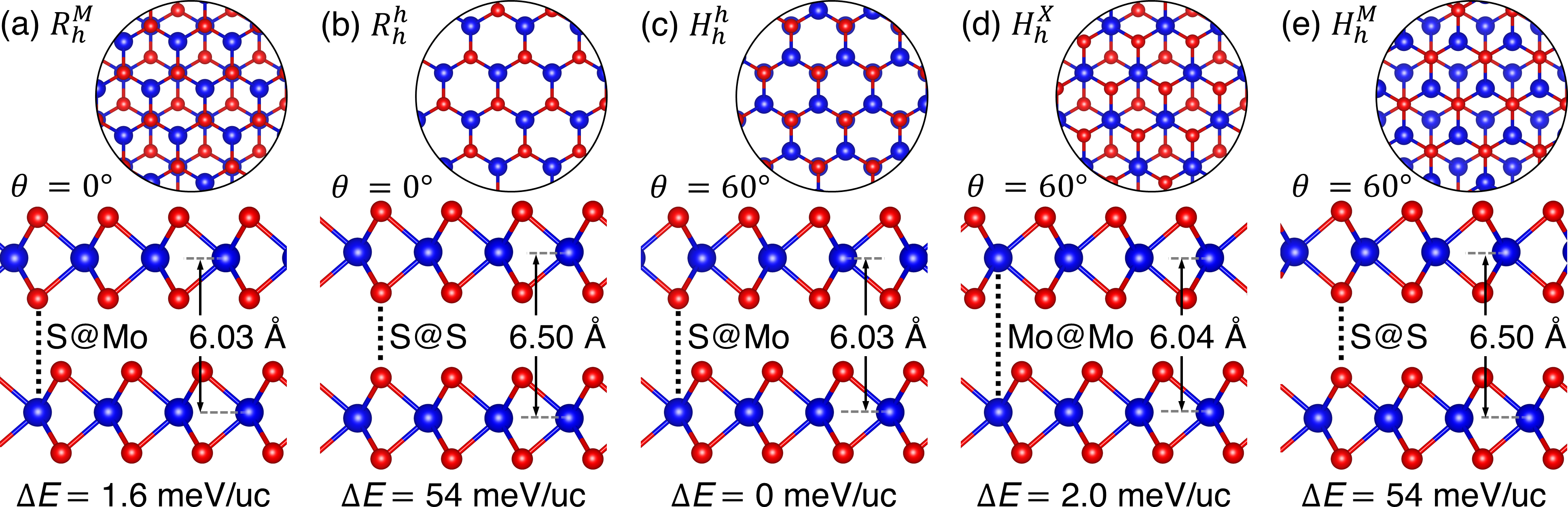}
    \caption{High-symmetry stackings of MoS\textsubscript{2} bilayers in the top and side view: (a) $R_h^M$ (corresponds to the 3$R$ stacking), (b) $R_h^h$, (c) $H_h^h$ (corresponds to the 2$H$ stacking), (d) $H_h^X$, and (e) $H_h^M$. The $R$-type corresponds to a $0^\circ$ rotation, the $H$-type to a $60^\circ$ rotation between the layers.~\cite{yu2018heterobilayers} The stacking type and the interlayer distance of the fully relaxed structures are given together with the total energy differences, $\Delta E$, relative to the lowest-energy stacking $H_h^h$. Energies are given in meV per unit cell (uc).}
    \label{fig_BL_stackings}
\end{figure*}

Initial tBL MoS\textsubscript{2} structures with periodic boundary conditions were constructed from ideal monolayer (ML) equilibrium geometries placed on top of one another with twist angles $\theta$ between $0^\circ$ and $60^\circ$.
Details on the structure generation are given in the Supporting Information (SI)~\cite{SI} (see Fig.~S2).
Full geometry optimization was performed employing the LAMMPS code~\cite{LAMMPS} using a reactive force field (ReaxFF)~\cite{ReaxFF1_vanDuin2001,ReaxFF2_chenoweth2008,ReaxFF3_aktulga2012} parametrized for MoS\textsubscript{2}.\cite{ostadhossein2017reaxff}
Geometry convergence was considered to be reached at a maximum force component of 0.1\,eV/\AA\ and the remaining pressure of the atomic system below 10 kbar.
We obtained 1219 fully optimized tBL structures with $\theta$ in the range of $0.2^\circ-59.6^\circ$.
All fully-relaxed tBL MoS\textsubscript{2} structures used in this work and the calculation files are available as ZENODO repository.\cite{ZENODO}
The results of the force field calculations were verified and are in good agreement with density-functional theory (DFT), as described in the SI (see Tables~S1, S2, and Fig.~S3).
While the interlayer distances in (t)BL structures optimized with ReaxFF and PBE+TS are in close agreement, we observe a consistent overestimation by about 0.35~\AA\ in PBE+MBD.
The moiré patterns introduced by rigidly twisted layers change with $\theta$, and the resulting size of the periodic unit strongly increases for small twist angles ($\theta \rightarrow 0^\circ$ and $\theta \rightarrow 60^\circ$).
While at $\theta=1.05^\circ$, there are already 17,862 atoms in the unit cell, this value reaches 493,026 atoms in the largest studied structure with $\theta=0.2^\circ$.

The primary descriptor for analysis of the structural properties is the local interlayer distance $d(x,y)$, defined as the vertical distance between the planes spanned by the Mo atoms of the two layers.
As the atoms in the two layers are not necessarily eclipsed, multivariate interpolation was applied to map the atomic coordinates to a quasi-continuous description of $d(x,y)$.
The size of the structural elements was determined using the so-called moiré cell -- the smallest repeating unit in the local interlayer distance landscape.
Additional information on the analysis of structural and energetic properties is given in the SI (see Figs.~S4 and S5).
A movie file showing how the interlayer distance landscape changes with $\theta$ is available as part of the ZENODO repository.\cite{ZENODO}

\begin{figure*}[ht!]
    \centering
    \includegraphics[width=\textwidth]{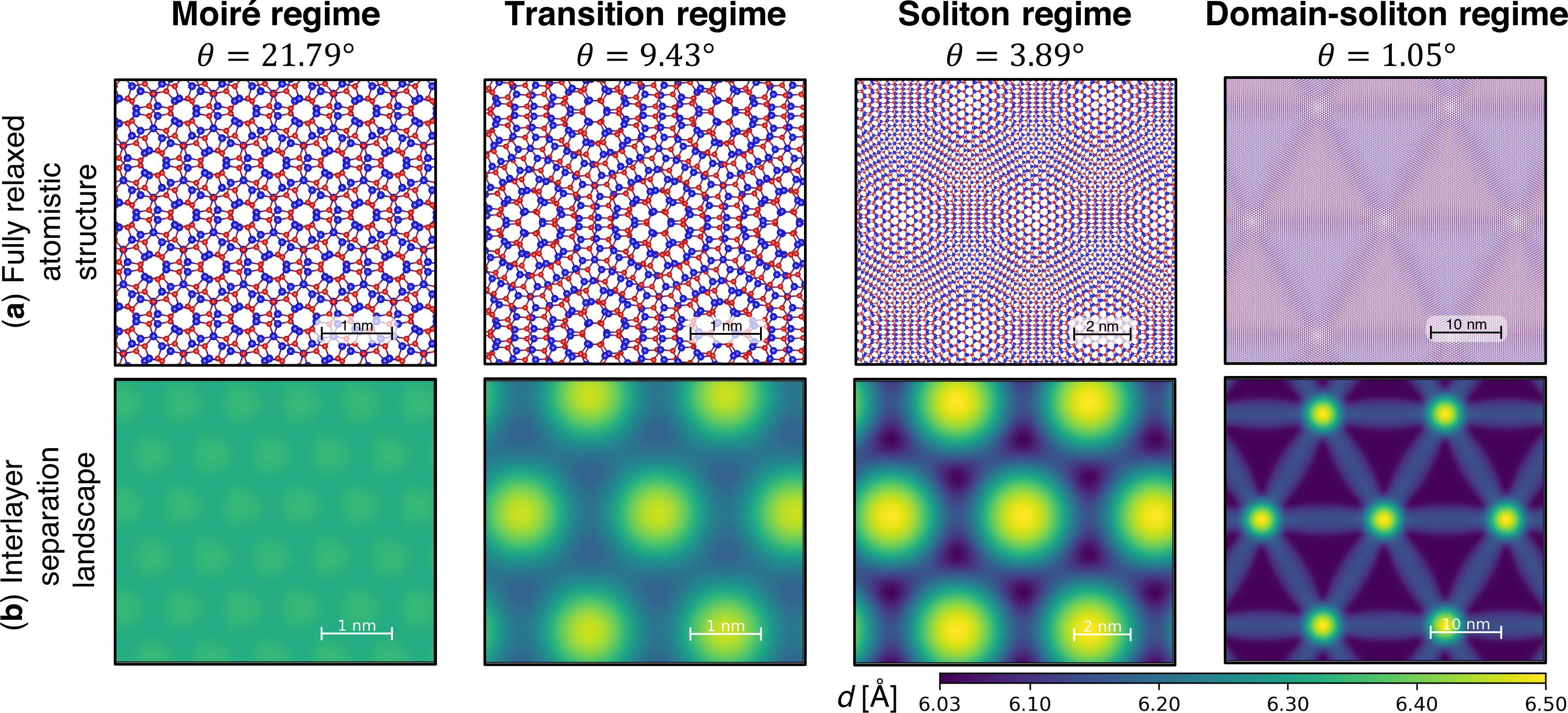}
    \caption{Atomistic structure of tBL MoS\textsubscript{2} systems with representative $\theta$ corresponding to the different structural regimes: moiré regime, transition regime, soliton regime, and domain-soliton regime. (a) Top views of the fully relaxed atomistic structures with Mo atoms in blue and S atoms in red. (b) Interlayer separation landscape, calculated from the positions of the Mo atoms.}
    \label{fig_regime_overview}
\end{figure*}

Electronic structure calculations were carried out using density functional theory (DFT) as implemented in FHI-aims.\cite{FHIaims_1,FHIaims_2,FHIaims_3}
The Perdew-Burke–Ernzerhof (PBE) exchange-corre\-la\-tion functional~\cite{PBE} was used together with a light tier 1 numeric atomic orbital basis set~\cite{blum2009nao_basis_fhiaims}, a converged Monkhorst-Pack $k$-point grid~\cite{monkhorst1976} with a density of 5~$\nicefrac{1}{\text{\AA}^{-1}}$ along the reciprocal lattice vectors, and non-self consistent spin-orbit coupling (SOC).\cite{FHIaims_4}
Using bilayer and twisted bilayer MoS\textsubscript{2} systems, we verified that band structures calculated using ReaxFF- and DFT-optimized structures are consistent (see Figs.~S6, S7, S8, and Table~S3).
Electronic properties were analyzed from band structure and density of state (DOS) for twist angles
between $3.89^\circ$ and $56.11^\circ$.
Atom-projected DOS was calculated with a Gaussian broadening of 50~meV.
From this, the local gap of the Mo atoms was approximated using the first maximum below the Fermi level as the valence band maximum (VBM) and the onset above the Fermi level as the conduction band minimum (CBM).
To further show which parts of the structure contribute to the VBM, the atom-projected DOS of the Mo atoms at the VBM energy of the whole structure was evaluated.

For exemplary systems, the electronic band structure and DOS were calculated using the lowest-order density-functional based tight-binding (DFTB) method, DFTB0,~\cite{porezag1995DFTB0} as implemented in QuantumATK,\cite{smidstrup2020quantumatk} employing the parameterization of Wahiduzzaman et al.~\cite{wahiduzzaman2013quasinano13}
The $k$-grid was set to $2\times2\times1$ for large supercells with $\theta \leq7^\circ$ and to $5\times5\times1$ for $\theta>7^\circ$.
The electronic structure calculated using DFTB was validated against DFT and found to be in fairly good agreement (see Figs.~S9, S10, and S11). 
We observe, however, a consistently larger band gap by $\approx0.4$~eV.

Ballistic electron transport calculations within DFTB were performed, employing the Landauer-Büttiker formalism and the non-equilibrium Green’s function (NEGF) approach.\cite{stradi2016general, PhysRevB.31.6207}
The device simulated with NEGF consists of three parts: the left and right (semi-periodic) leads attached to the finite central scattering region (see Fig.~S12 for an exemplary device configuration).
A dense $k$-point sampling of $1\times5\times99$ was used.
The whole device is periodic normal to the transport direction in the xy plane.
A vacuum of 20~\AA\ normal to the bilayer plane was used to avoid any interaction caused by periodic images.

Details for all methods used in this work are given in the SI.

\section{Results and Discussion}

\subsection{Structural and energetic properties}

\begin{figure*}[ht!]
    \centering
    \includegraphics[width=\textwidth]{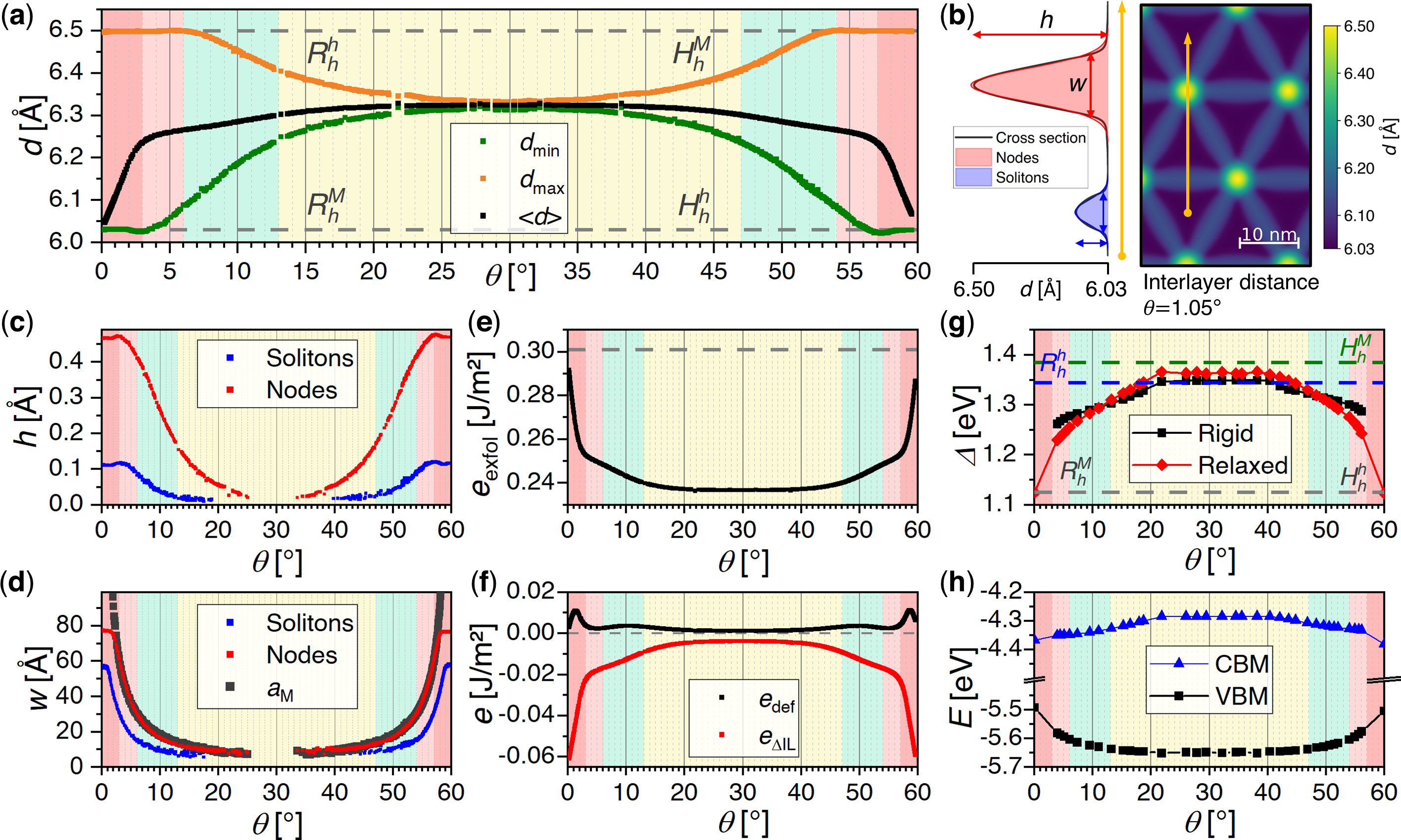}
    \caption{Structural, energetic, and electronic properties of tBL MoS\textsubscript{2} as function of $\theta$. (a) Maximum, minimum, and mean interlayer distance, $d$, calculated from the Mo atoms in the neighbouring layers. The reference lines correspond to the interlayer distance in the high-symmetry BL stackings. (b) Cross-section through the interlayer distance landscape for $\theta=1.05^\circ$. A Gaussian is fitted to the node (red) and soliton (blue) to approximate their width, $w$.
    (c) Numerically determined height, $h$, of solitons (blue) and nodes (red), obtained from the maximum interlayer distance inside the structural element relative to the minimum value of the structure.
    (d) Width, $w$, of solitons (blue) and nodes (red), obtained from the base width $4\sigma$ of the Gaussian fitted to the cross-section as shown in (b). The numerically determined moiré periodicity, $a_\mathrm{M}$, is shown in grey. (e) Exfoliation energy of tBL MoS\textsubscript{2}. The $R_h^M$ and $H_h^h$ stackings with the highest value of 0.30~J/m² are indicated with a dashed blue line. (f) Contributions to the exfoliation energy originating from the deformation of the layers and change of the interlayer adhesion due to the corrugation. (g) Band gaps for rigidly twisted and fully relaxed structures. As a reference, the band gap of the relaxed BL systems is shown. The relaxed ML has a band gap of 1.65~eV. (h) Absolute energy of VBM and CBM when including vacuum corrections. In all subplots, the structural regimes are indicated with the (domain-)soliton regime in (dark) light red, the transition regime in green, and the moiré regime in yellow.}
    \label{fig_theta_dependence}
\end{figure*}

The primary reference systems needed for the analysis of tBL MoS\textsubscript{2} are the high-symmetry stacking configurations in untwisted bilayer (BL) MoS\textsubscript{2}, as shown in Fig.~\ref{fig_BL_stackings}.
The calculated structural properties of the BL MoS\textsubscript{2} stackings are consistent with previous reports in the literature~\cite{he2014stacking} with more details given in the SI (see Table~S4).
In contrast to heterobilayer systems, there are only five possible stackings because $R_h^M$ and $R_h^X$ are equivalent in homobilayer systems.
$R_h^M$ and $H_h^h$ belong to the S@Mo (read "S on Mo") stacking type -- low-energy stackings with the smallest interlayer distance of 6.03~\AA, where the S atoms in the top layer eclipse the Mo atoms in the bottom layer.
$R_h^h$ and $H_h^M$ belong to the S@S stacking type -- high-energy stackings which are less stable by 54~meV/uc and have a large interlayer distance of 6.50~\AA.
$H_h^X$ belongs to the Mo@Mo stacking type with energy and interlayer distance comparable to that of $H_h^h$.
With this, three high-symmetry stacking configurations, namely $R_h^M$, $H_h^h$, and $H_h^X$, with low energy and short interlayer distances exist.
Their relative energies are within 2~meV per primitive BL unit cell (uc), consisting of 2 MoS\textsubscript{2} formula units.
This results in the stacking domains for $\theta\rightarrow0^\circ$ showing local $R_h^M$ stacking, while for $\theta\rightarrow60^\circ$ they alternate between $H_h^h$ and $H_h^X$.
This makes the $R$-type stacking domains energetically equivalent, while the $H$-type stacking domains have different stacking energy, resulting in a size difference between adjacent stacking domains for twist angles near $60^\circ$ (see Fig.~S1).~\cite{arnold2023chemRxiv}
Local $R_h^h$ ($H_h^M$) stacking arrangements are present within the nodes for $\theta\rightarrow0^\circ$ ($\theta\rightarrow60^\circ$).

\begin{figure*}[ht!]
    \centering
    \includegraphics[width=\textwidth]{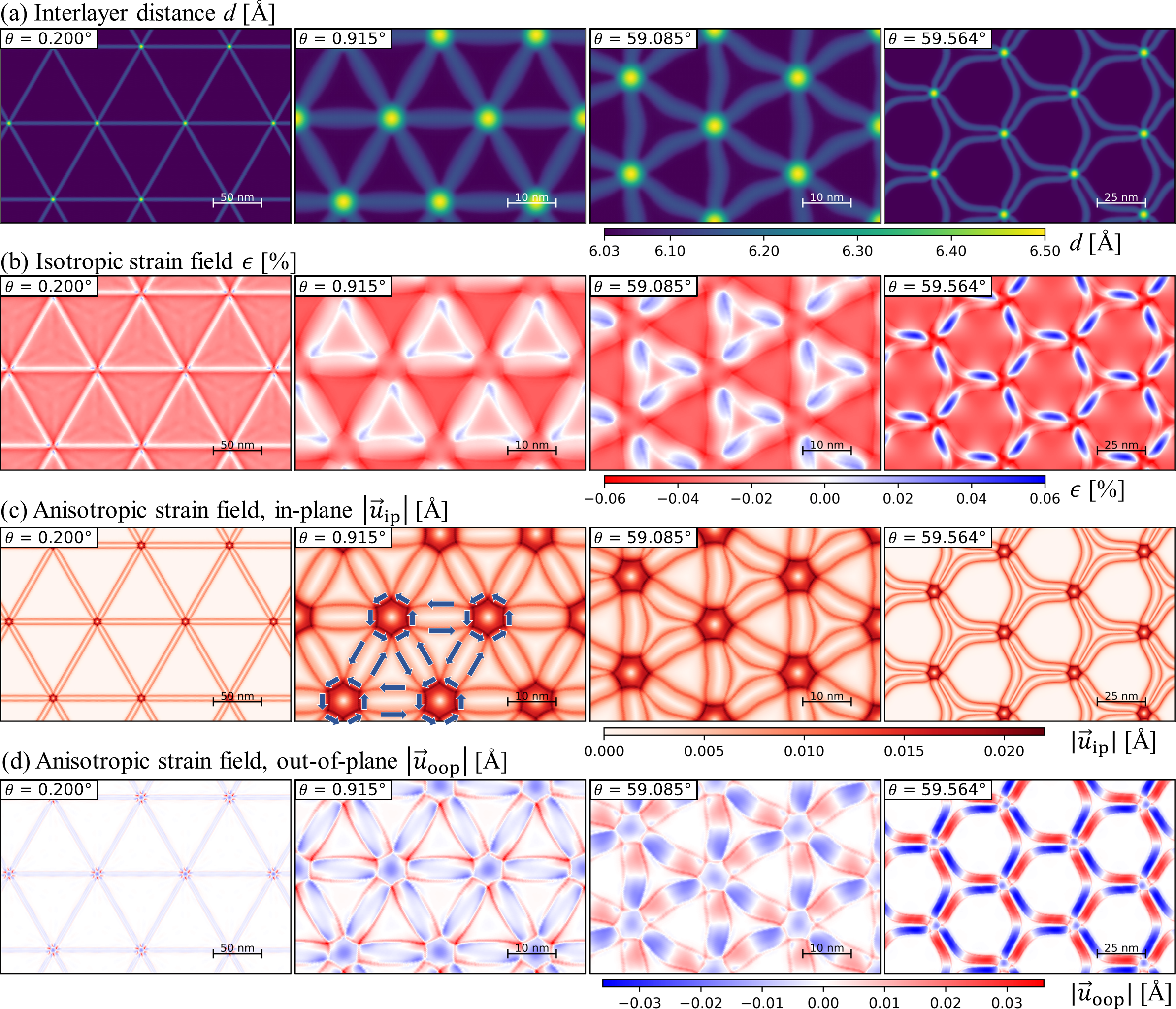}
    \caption{Selected structural properties of exemplary tBL MoS\textsubscript{2}: (a) Local interlayer distance $d$. (b) Isotropic strain field, $\epsilon$, relative to the unstrained MoS\textsubscript{2} ML. Areas of compressive strain are shown in red, tensile strain in blue. (c) Absolute value of the in-plane component of the anisotropic strain field, $|\vec{u}_{\mathrm{ip}}|$, with the qualitative directional component of the vector field indicated for $\theta=0.915^\circ$ using arrows. (d) Out-of-plane component of the anisotropic strain field, $|\vec{u}_{\mathrm{oop}}|$. The properties in (b-d) are plotted only for the upper layer and interpolated between the data points.}
    \label{fig_strain_field_combined}
\end{figure*}

Structural and electronic properties of tBL MoS\textsubscript{2} gradually change as a function of $\theta$.
As most properties of tBL MoS\textsubscript{2} show a continuous change with $\theta$, there is a continuous transition between the soliton and moiré regimes rather than a single "magic" twist angle at which this transition occurs.
Nevertheless, for certain $\theta$ particular structural transitions emerge.
Here, we provide a comprehensive characterization of structural properties as a function of $\theta$.

Exemplary structures for each regime are given in Fig.~\ref{fig_regime_overview} and details on the criteria used for the in-depth description of the regimes are given in the SI.
The moiré regime is present for large $\theta$ and can be modeled with rigidly twisted layers.
When going towards smaller $\theta$, the characteristics of the system change as the influence of relaxation becomes more prominent.
The structural elements start to appear, indicating the beginning transition towards the soliton regime.
These intermediate twist angles can be classified by a transition regime.
This accounts for the continuous transition towards the (strain) soliton regime.\cite{yoo2019graphene,kumar2015elastic,naik2019KCSW}
In this work, it is divided into two parts due to structural and electronic property differences:
For small twist angles, we classify the soliton regime.
It exhibits all characteristic structural elements:
The stacking domains form a honeycomb lattice but are still relatively small.
The solitons are present as bridges between the nearby nodes.
When going towards the smallest $\theta$, a more extensive atomic reconstruction is observed:
Extended stacking domains dominate the structure.
The solitons form a hexagonal network with the nodes positioned at the intersection points.
To account for these structural differences, We classify the smallest twist angles as the domain-soliton regime.

The structural and energetic properties studied as a function of $\theta$ are:
(i) The interlayer distance $d$ with its minimum, $d_\mathrm{min}$, maximum, $d_\mathrm{max}$, and mean, $d_\mathrm{mean}$, values as shown in Fig.~\ref{fig_theta_dependence}(a).
Here, gaps in the curves around $\theta=21.79^\circ$, $38.21^\circ$, and $13.17^\circ$ are visible.
They originate from "magic angles" with respect to the size of the moiré cell.
A small cell can be realized at these values while diverging to unfeasibly large structures at nearby $\theta$.
(ii) The spatial extension of nodes and solitons, analyzed by fitting a Gaussian function to their cross-section in the interlayer distance landscape as demonstrated in Fig.~\ref{fig_theta_dependence}(b).
Their height, $h$, as the maximum corrugation of the interlayer distance inside a structural element, is visualized in Fig.~\ref{fig_theta_dependence}(c).
Their width, $w$, is shown in Fig.~\ref{fig_theta_dependence}(d), using the diameter of the circular nodes and the width at the center of the solitons to approximate their spatial extension.
This property is also compared to the moiré periodicity (see Figs.~S13 and S14 for more details) as the periodic unit of the moiré pattern and used to estimate the area of the stacking domains (see Fig.~S15).
(iii) The exfoliation energy is given in Fig.~\ref{fig_theta_dependence}(e).
It is further separated into individual contributions from the deformation of the individual layers and the change of the interlayer adhesion (see Fig.~\ref{fig_theta_dependence}(f)) due to the relaxation of the layers (for more details see Fig.~S16).
The exfoliation energy as a function of $\theta$ can be described by piecewise linear parts (see Fig.~S17 and Table~S5).

Additionally, the displacement and strain fields were analyzed.
The displacement field (see Fig.~S18) allows to gain insight into the atomic reconstruction during relaxation.\cite{nam2017lattice}
The intralayer strain distribution was investigated using the isotropic and anisotropic strain fields, as shown in Fig.~\ref{fig_strain_field_combined}, both giving access to different structural information.\cite{vanWijk2015}
While the scalar isotropic strain field, $\epsilon$, gives the local strain relative to an unstrained ML, the anisotropic strain field, $\vec{u}$, shows the directional change of the strain as a vector field.
Equations and additional information are also given in the SI (see Fig.~S19).

Inside the moiré regime, between $\approx 13^\circ$ and $\approx47^\circ$, an almost constant value of $d_\mathrm{mean}$ with only minor interlayer distance modulation is found (see Fig.~\ref{fig_theta_dependence}(a)).
This indicates that the rigidly twisted model can be applied in this regime, and the structural elements are not pronounced.
Similarly to the interlayer distance, the exfoliation energy is almost constant with a minor change of $\approx 0.15$~mJ/m² per $1^\circ$.
The anisotropic strain field remains non-zero, as the anisotropy becomes zero only for untwisted bilayers at $0^\circ$ and $60^\circ$ (see Fig.~S19).
The displacement field being close to zero confirms that the ip reconstruction is negligible (see figure S18(b)).

In the transition regime, $6^\circ<\theta<13^\circ$ (equivalently, $47^\circ<\theta<54^\circ$), the atomic reconstruction becomes more prominent.
Here, the interlayer distance shows a slow decrease of $d_\mathrm{mean}$ with decreasing $\theta$.
This is reflected in the exfoliation energy, which slowly increases by $\approx 1.4$~mJ/m² per $1^\circ$.
A strongly increasing interval $\Delta d=d_\mathrm{max}-d_\mathrm{min}$ (from 0.16~\AA\ at $13^\circ$ to 0.41~\AA\ at $6^\circ$) is found, showing an intensifying oop deformation of the individual layers (see Fig.~\ref{fig_theta_dependence}(a)).
Contrary to this, the ip displacement field still shows a negligible magnitude (see Fig.~S18(b)) with mean values below 0.03~\AA.
This confirms oop relaxation as the dominating effect of the atomic reconstruction in this regime, associated with emerging structural elements.
An increasing spatial extension of nodes and solitons with decreasing $\theta$ is observed as a strong increase in their height and width, as shown in Fig.~\ref{fig_theta_dependence}(c, d).
While at $13^\circ$, the base width of the nodes (solitons) is 13~\AA\ (8~\AA), it increases to 30~\AA\ (15~\AA) at $6^\circ$.
The nodes are the dominating structural element of the transition regime as they appear already at larger values of $\theta$ near $13^\circ$ ($47^\circ$).
Solitons appear only for lower $\theta$ near $6^\circ$ ($54^\circ$), indicating the beginning of the soliton regimes.
At the same time, the anisotropy of the strain field is becoming large and shows an emerging order with decreasing $\theta$ (see figure S19).

In the soliton regime, $3^\circ<\theta\leq6^\circ$ ($54^\circ\leq\theta <57^\circ$), the atomic reconstruction continues to become more prominent, and all three structural elements are present.
The $d_\mathrm{max}$ value reaches its upper bound, equivalent to $R_h^h$ ($H_h^M$) stacking (S@S type), at $\theta=6^\circ$ ($\theta=54^\circ$).
This correlates with fully pronounced nodes of local high-energy stacking type.
The height of the structural elements increases and reaches its maximum value at $\theta=3^\circ$ ($\theta=57^\circ$).
Similarly, the width of nodes and solitons also increases.
For $\theta$ down to $\approx3^\circ$ ($\approx57^\circ$), the width of the nodes coincides with $a_\mathrm{M}$, showing that the nodes are the dominating structural element of the moiré cell.
Their large size also results in the solitons being present only as short bridges between adjacent nodes, as shown in Fig.~\ref{fig_regime_overview}(b).
The stacking domains are still small but already visible in the interlayer distance landscape, forming a honeycomb lattice.
They exhibit their characteristic local S@Mo stacking, but the interlayer distance of the S@Mo BL is not yet reached, as visible from $d_\mathrm{min}$ decreasing further (see figure \ref{fig_theta_dependence}(a)).
Their small size is also visible from $d_\mathrm{mean}$ and the exfoliation energy, which show a small change with decreasing $\theta$.
The dominating factor in the atomic reconstruction is still the oop relaxation.
The displacement field becomes more sizeable in this regime with values up to 0.08\,\AA\ (see Fig.~S18), showing the beginning influence of the ip reconstruction.

The final regime is the domain-soliton regime, $0^\circ<\theta\leq3^\circ$ ($57^\circ\leq\theta < 60^\circ $).
Additionally to the oop relaxation, a significant ip reconstruction is present in this regime.
This can be deduced from the displacement field, which shows strongly increasing values of up to $\approx$0.8~\AA\ (see Fig.~S18(b)).
Its largest values are at the boundary between the structural elements.
The center of nodes and stacking domains form high-symmetry stackings in the rigidly twisted structure and, thus, can be understood as seeds for the atomic reconstruction.

In the domain-soliton regime, $d_\mathrm{min}$ reaches its lower bound, equivalent to $R_h^M$ and $H_h^h$ stackings, showing that the stacking domains are fully pronounced.
As $\Delta d$ becomes maximal, the height of nodes and solitons reaches a plateau, with the nodes being at least three times larger than the solitons (see Fig.~\ref{fig_theta_dependence}(c)).
A plateau is also observed for their width at $\theta\leq1^\circ$ ($\theta\geq59^\circ$): the solitons reach a maximum of 57~\AA, and the nodes a value of 77~\AA\ (see Fig.~\ref{fig_theta_dependence}(d)).
In the literature, the appearance of such a plateau is also described.\cite{carr2018relaxation}
Naik et al.\cite{naik2018ultraflatbands} find the full width at half maximum (FWHM) of the solitons to be $\approx$15~\AA\ at $3.48^\circ$.
This is in excellent agreement with our results, giving $w=$22.7~\AA\ for $3.48^\circ$, corresponding to a FWHM of 15.7~\AA.
As the spatial extension of the nodes becomes maximal, their width no longer coincides with the moiré periodicity $a_\mathrm{M}$, as shown in Fig.~\ref{fig_theta_dependence}(d).
Decreasing $\theta$ increases the separation between the nodes, creating a pronounced network of solitons and extended stacking domains in the domain-soliton regime (see Fig.~\ref{fig_regime_overview}(b)), giving the regime its name.
The stacking domains become the dominating structural element (see Fig.~S15).
At $3^\circ$, only 1.5\% of the moiré cell can be accounted for the stacking domains, compared to up to 80\% at $0.2^\circ$.
The formation of extended stacking domains is reflected in $d_\mathrm{mean}$ strongly decreasing towards the value of the S@Mo stacking type (see Fig.~\ref{fig_theta_dependence}(a)).
It is also visible as a strong increase of the exfoliation energy by $\approx21$~mJ/m² per $1^\circ$ (see Fig.~S17 and Table~S5).
Strain effects account for only $\approx$0.1\% of the exfoliation energy, the deformation of individual layers $\approx$1\%, making the interlayer adhesion terms the dominating contributions (see Fig.~S16).
The values of the exfoliation energy (0.29~J/m² for small $\theta$ and 0.24~J/m² for large $\theta$) are in good agreement with results by Emrem et al.,\cite{emrem2022london} who reported 0.33~J/m² for BL MoS\textsubscript{2}, calculated at random-phase approximation (RPA) level.
The structural properties described for the domain-soliton regime are reflected in the strain fields, as shown for examples with $\theta \rightarrow 0^\circ$ and $\theta \rightarrow 60^\circ$ in Fig.~\ref{fig_strain_field_combined}.
The stacking domains exhibit a constant isotropic strain $\epsilon$ with negligible anisotropy both ip and oop, confirming that these are large, uniform S@Mo areas.
The nodes and boundary region between solitons and stacking domains, on the other hand, show strong anisotropy.
As indicated in Fig.~\ref{fig_strain_field_combined}(c), the directional component has a circular arrangement around the nodes and, as reported by Alden et al., shows the presence of sheer strain at the solitons.\cite{alden2013solitons}
Due to this strain accumulation, the solitons are often called strain solitons.
The oop component, qualitatively comparable to the curvature of the individual layers, points in opposite directions in the individual layers.
It is largest in nodes and solitons, as the main curvature of the layers is localized in these structural elements.

It is important to note that the interlayer separation landscape in structures with $\theta\rightarrow0^\circ$ and $\theta\rightarrow60^\circ$ deviates strongly for $\theta<1^\circ$ and $\theta>59^\circ$ (see Fig.~\ref{fig_strain_field_combined}(a)).
For $\theta\rightarrow0^\circ$, the stacking domains are equivalent, and the solitons are linear.
This changes for $\theta\rightarrow60^\circ$, where small and large stacking domains alternate, separated by bent solitons.
Different symmetries cause this: in $\theta\rightarrow0^\circ$, all stacking domains exhibit a local $R_h^M$ stacking, while in $\theta\rightarrow60^\circ$, the adjacent domains alternate between $H_h^h$ and $H_h^X$ stackings, as shown in Fig.~S1.\cite{carr2018relaxation,maity2020phononsTMDCs,arnold2023chemRxiv}
The $H_h^h$ and $H_h^X$ stackings differ in energy (2\,meV/uc, see Fig.~\ref{fig_BL_stackings}).
Thus, the size of stacking domains also differs, with the regions of more stable stacking becoming larger.
This results in the formation of bent solitons and a lowering of the overall symmetry.
It is accompanied by structures with $\theta\rightarrow60^\circ$ being stronger bound by up to 1.6\,mJ/m² compared to $\theta\rightarrow 0^\circ$ (see Fig.~S20), resulting from a stronger adhesion between the layers.
We observe that even though there are such substantial differences between $\theta\rightarrow0^\circ$ and $\theta\rightarrow60^\circ$, there is an almost ideal symmetry $\theta_i=60^\circ-\theta'_i$ for the twist angles at which the transition between the regimes takes place.
It is interesting to note that introducing an inherent interlayer strain also results in a bending of the solitons, as shown in the SI (see Fig.~S21).

Our characterization approach is fully consistent with the literature results and shares common criteria other authors used to asses the transition between the regimes.
Garguilo et al.\cite{gargiulo2017graphene} analyze the twist energy -- a property comparable to the deformation energy used here -- of tBL graphene as a function of $a_{\mathrm{M}}$.
They find $1.2^\circ$ as the crossover angle between a "solitonic regime" and a "rigid regime."
Dai et al.\cite{dai2016twistedgraphene} describe the transition between the regimes to happen at the coincidence point between $a_\mathrm{M}$ and the width of the solitons.
They find values from $1^\circ$ to $1.6^\circ$ for tBL graphene, similar to the value of $3^\circ$ in the more rigid MoS\textsubscript{2} systems.\cite{elder2015bending_graphene_vs_mos2}
Maity et al.\cite{maity2020phononsTMDCs} find a maximum deformation of the individual layers in tBL MoS\textsubscript{2} at $3^\circ$ and a transition angle of $13^\circ$.
This is in direct agreement with the begin of the domain-soliton regime and the moiré regime as defined in this work.
Wang et al.\cite{wang2020tmdcs} describe the transition between the structural regimes to take place in an interval from $4^\circ$ to $5.1^\circ$ for tBL WSe\textsubscript{2}. 
This agrees with the soliton regime, where the structural elements start to vanish.
Enaldiev et al.\cite{enaldiev2020stacking} find values of $2.5^\circ$ and $59^\circ$ in tBL TMDC systems as characteristic twist angles for forming extended stacking domains.
This is the only other work known by the authors to explicitly consider the range $30^\circ\leq\theta<60^\circ$ and agrees with the newly defined domain-soliton regime.
Overall, we find an excellent agreement of the characterization derived in this work with previously published results.
Our analyses are based on a much larger set of twisted bilayer structures, allowing us to find a smooth transition between the regimes.
The approach proposed in this work is expected to be transferable to different twisted homobilayers and, after small adaptations, to other systems, including heterobilayers.

\subsection{Electronic properties}

\begin{figure*}[ht!]
    \centering
    \includegraphics[width=\textwidth]{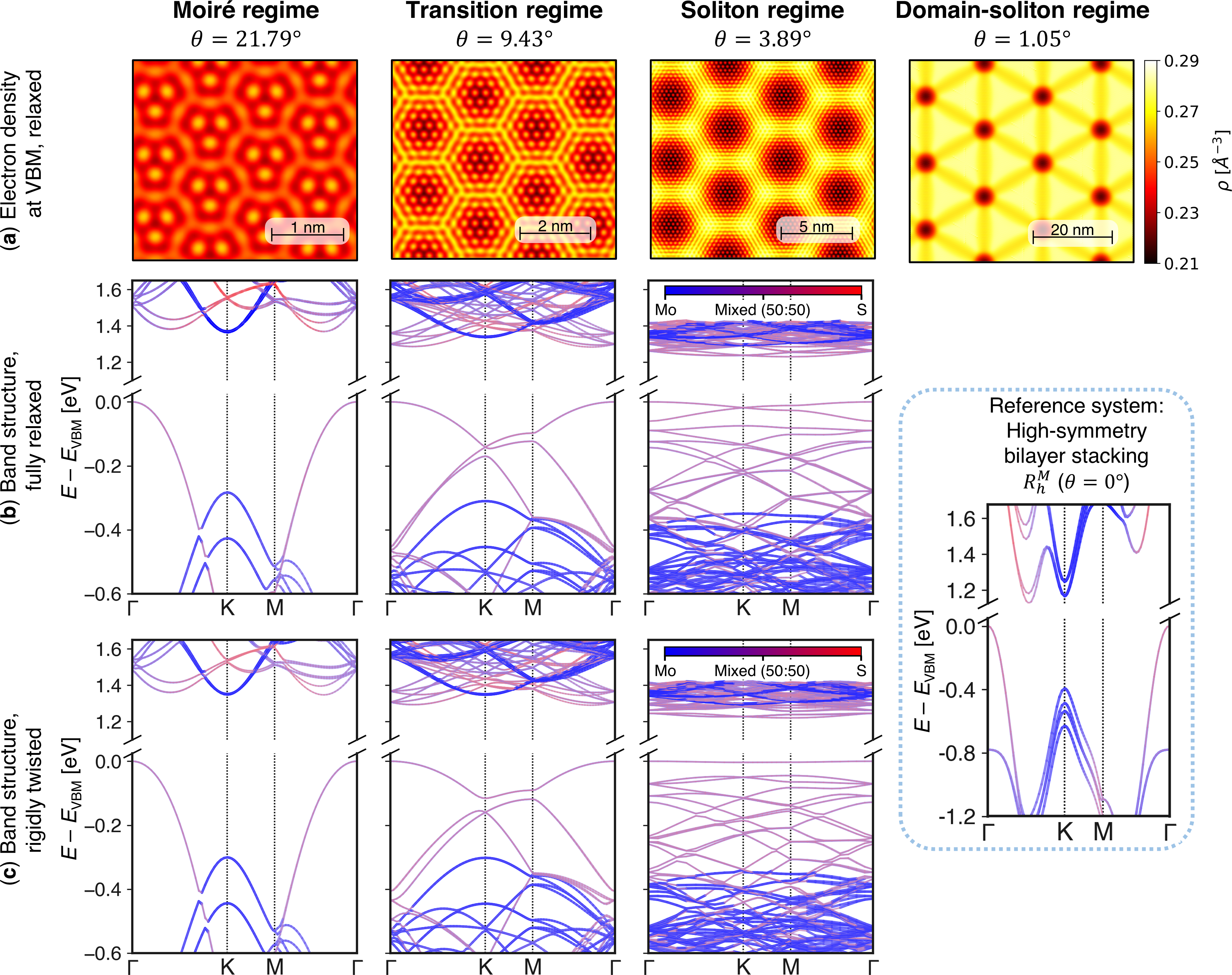}
    \caption{(a) Electron density at the VBM, calculated at DFTB level, for representative structures in the structural regimes, corresponding to Fig.\ref{fig_regime_overview}. (b, c) Atom-projected band structures of the (b) fully relaxed and (c) rigidly twisted systems, including the $R_h^M$ bilayer as a reference.}
    \label{fig_electronics_by_regime}
\end{figure*}

We now investigate the impact of structural relaxation on the global and local electronic properties of tBL MoS\textsubscript{2} and how they change in the different regimes.
As a starting point, the electron density at the VBM is shown in Fig.~\ref{fig_electronics_by_regime} for examples of the different regimes.
We observe that it follows qualitatively the behavior of the interlayer distance landscape.
Only small modulations are found within the moiré regime, which increase in the transition regime.
In the soliton regime, the nodes become pronounced as areas of minimum electron density.
In contrast, extended stacking domains with uniform maximum electron density and a network of solitons are found in the domain-soliton regime.

Changes between the regimes are also found in the atom-projected band structures, as shown in Fig.~\ref{fig_electronics_by_regime}(b, c).
All tBL MoS\textsubscript{2} systems are indirect band gap semiconductors, with the VBM at $\Gamma$, similar to the untwisted bilayer limits.
The CBM is positioned at $K$, which changes to be along the path between $M$ and $\Gamma$ with decreasing $\theta$ in the transition regime.
An exception are those systems with a direct gap due to backfolding (see SI, Table~S3; unfolding these band structures is beyond this work's scope).
The atomic projection reveals that the frontier bands have a strongly mixed character (contributions from both the Mo and S atoms).
These bands show the most substantial changes between rigid and relaxed models.
Occupied bands with an increased Mo character are away from the frontier bands.
They show only minor relaxation effects but are more sensitive against the method used for geometry optimization.

The shape of bands and the band gap values change with $\theta$ (see Fig.~\ref{fig_theta_dependence}(g)).
The band gap difference between fully relaxed and rigidly twisted systems is very small, with a maximum of 44~meV, showing that there is just a minor influence of relaxation effects on this property.
At large $\theta$ in the moiré regime, the band structure differences between relaxed (Fig.~\ref{fig_electronics_by_regime}(b)) and rigidly twisted (Fig.~\ref{fig_electronics_by_regime}(c)) structures are also minor.
The band gap does not change with $\theta$ around $30^\circ$.
This further supports the validity of the model of rigidly twisted layers in this regime.
With decreasing $\theta$, the band gap slowly decreases and approaches the value of the S@Mo BL system, showing a linear dependence on $d_\mathrm{mean}$ (see SI for details).
This change originates from an increase of the VBM and a decrease of the CBM (see Fig.~\ref{fig_theta_dependence}(h)), with the former being the dominating factor.
Furthermore, both valence and conduction bands show a decreasing dispersion.
This trend continues in the soliton regime.

\begin{figure}[ht!]
    \centering
    \includegraphics[width=0.48\textwidth]{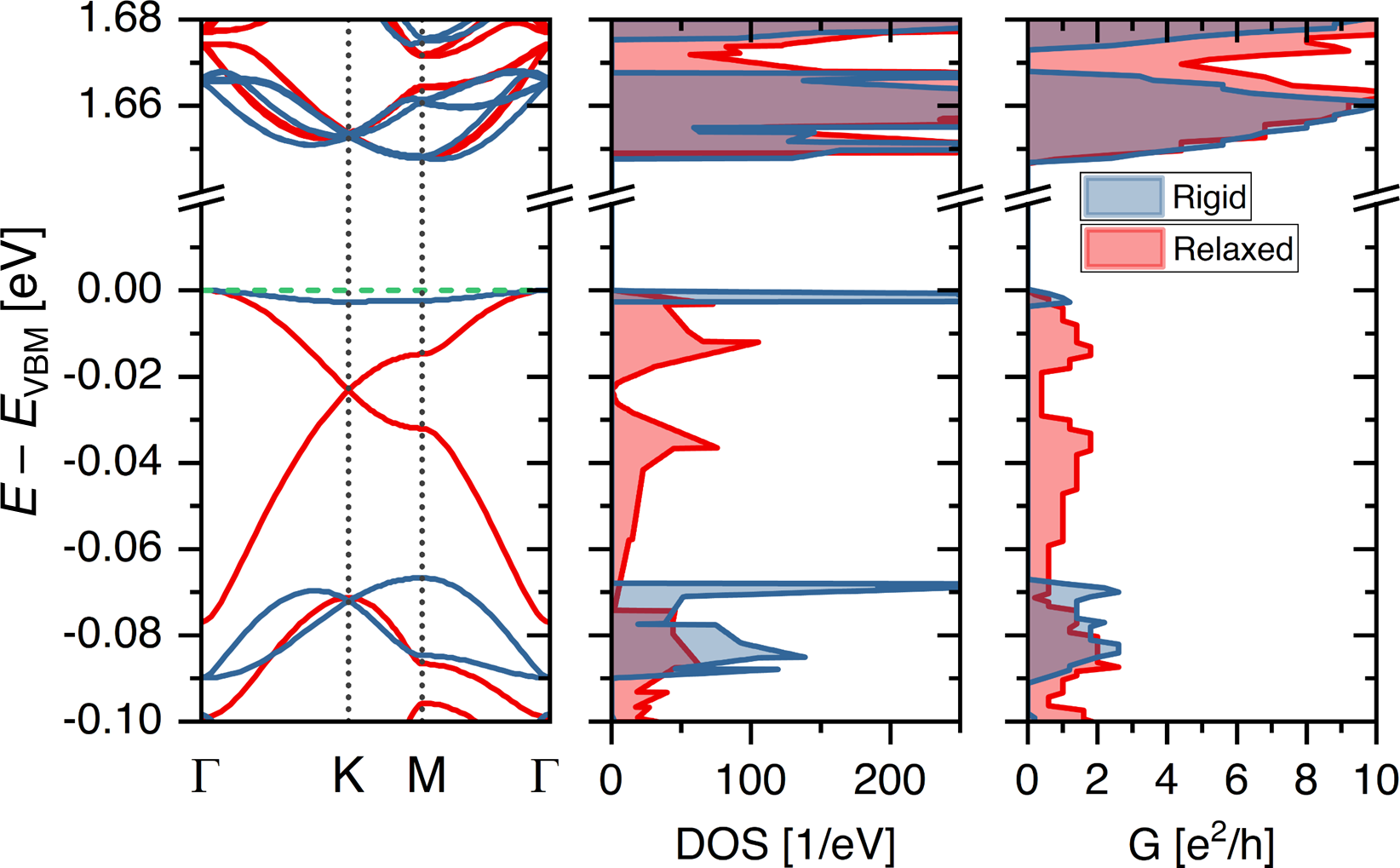}
    \caption{Electronic band structure, the corresponding DOS and conductance for both the rigidly twisted (blue) and fully relaxed (red) structure for tBL MoS\textsubscript{2} with $\theta=3.89^\circ$, computed at the DFTB level. The device configuration consists of left and right semi-infinite leads and a finite central region of three unit cells (see Fig.~S12) to ensure a smooth transition between the potential of the central region and the leads.}
    \label{fig_DFTB_03.89deg}
\end{figure}

Fig.~\ref{fig_DFTB_03.89deg} shows an exemplary band structure of tBL MoS\textsubscript{2} with $\theta=3.89^\circ$, calculated for relaxed and rigidly twisted cases.
The band gap values of both systems are about 1.25~eV.
However, the electronic structures differ significantly.

The top VB becomes flat in the rigidly twisted system, giving rise to van Hove singularity in the DOS.
Below that flat band, there is a gap of about 60~meV.
In the fully relaxed system, the corrugation of the layers results in the formation of a Dirac cone at $K$ close to the Fermi level.~\cite{arnold2023chemRxiv}
This can also be seen as a parabola-like shape of the DOS, similar to graphene.
Below the Dirac bands, characteristic kagome bands are observed.
Compared to the rigidly twisted system, the stronger band dispersion flattens the DOS at the VBM and results in a higher conductance in the relaxed system.

\begin{figure}[ht!]
    \centering
    \includegraphics[width=0.48\textwidth]{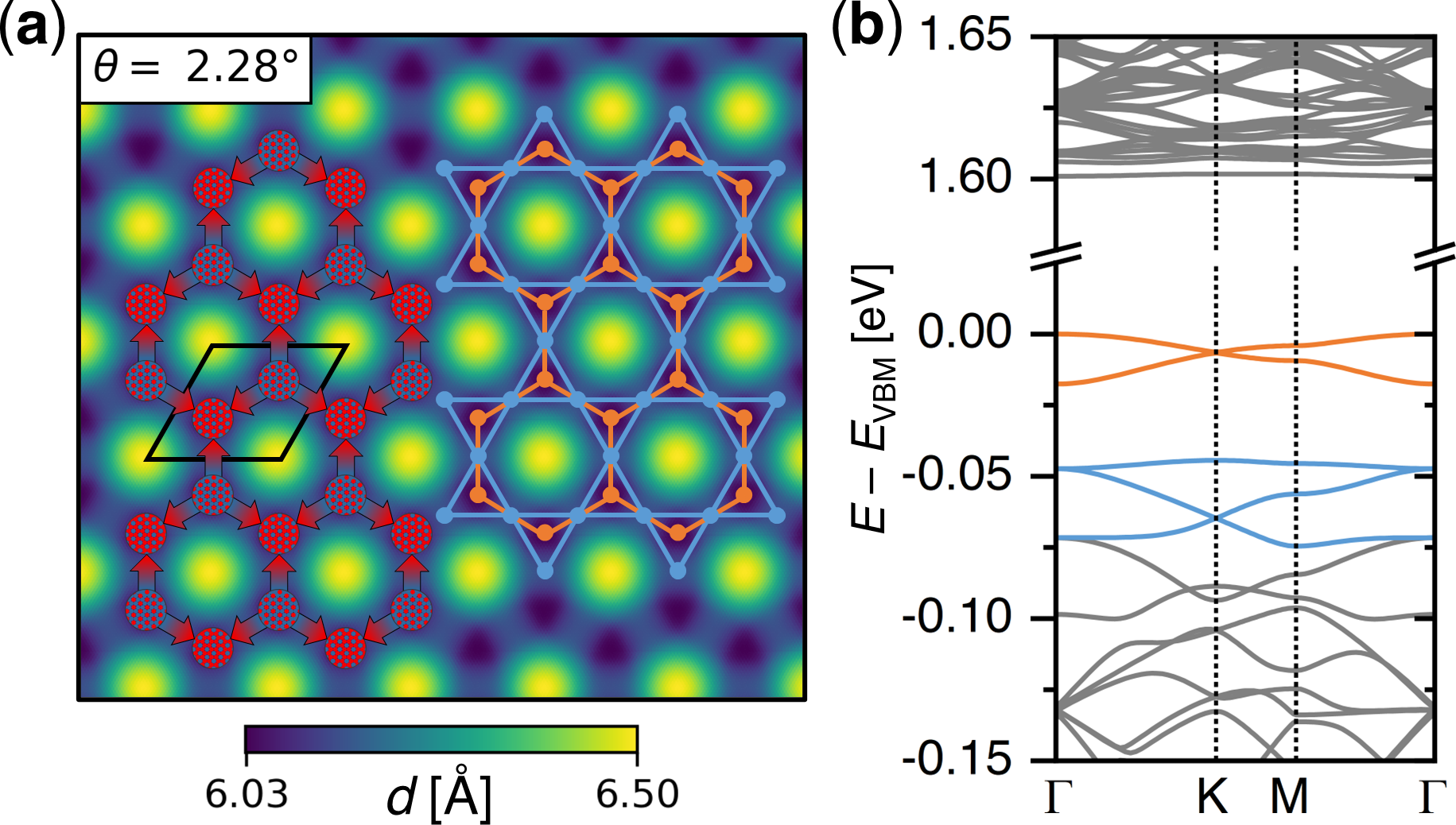}
    \caption{(a) Interlayer separation landscape with a schematic visualization of (left overlay) the structural elements and (right overlay) the superlattices formed by the structural elements in fully relaxed tBL MoS\textsubscript{2} with $\theta=2.28^\circ$. The stacking domains form a honeycomb lattice (orange) with two sites per unit cell, while the solitons form a kagome lattice (blue) with three sites per unit cell. (b) Electronic band structure of fully relaxed tBL MoS\textsubscript{2} with $\theta=2.28^\circ$, calculated at the DFTB level. The bands corresponding to the honeycomb and kagome lattice are highlighted.}
    \label{fig_superlattice_bands}
\end{figure}

The emergence of Dirac bands in the fully relaxed system is related to the formation of a honeycomb lattice of strongly reconstructed stacking domains in the (domain-)soliton regime (see Fig.~\ref{fig_superlattice_bands}(a)).
It is also described in several other publications.\cite{naik2020origin,angeli2021macdonald,xian2021pzbands,zhang2021pzbands,arnold2023chemRxiv}
It can further be seen that the three VBs below the Dirac bands form distorted kagome-type dispersion~\cite{springer2020topological} (blue in Fig.~\ref{fig_superlattice_bands}(b)), originating from a kagome lattice formed by the centers of the solitons (see Fig.~\ref{fig_superlattice_bands}(a)).
The formation of Dirac bands due to the honeycomb superlattice of the stacking domains can be observed both in the soliton and the domain-soliton regime.
It becomes more pronounced as $\theta$ approaches $0^\circ$, as shown in Fig.~S22(a).
Their bandwidth strongly decreases with decreasing $\theta$ (see Fig.~S22(b)) and is expected to result in flat bands for the smallest $\theta$.
We discuss these superlattice effects in more detail in another publication.\cite{arnold2023chemRxiv}

\begin{figure*}[htp]
    \centering
    \includegraphics[width=0.75\textwidth]{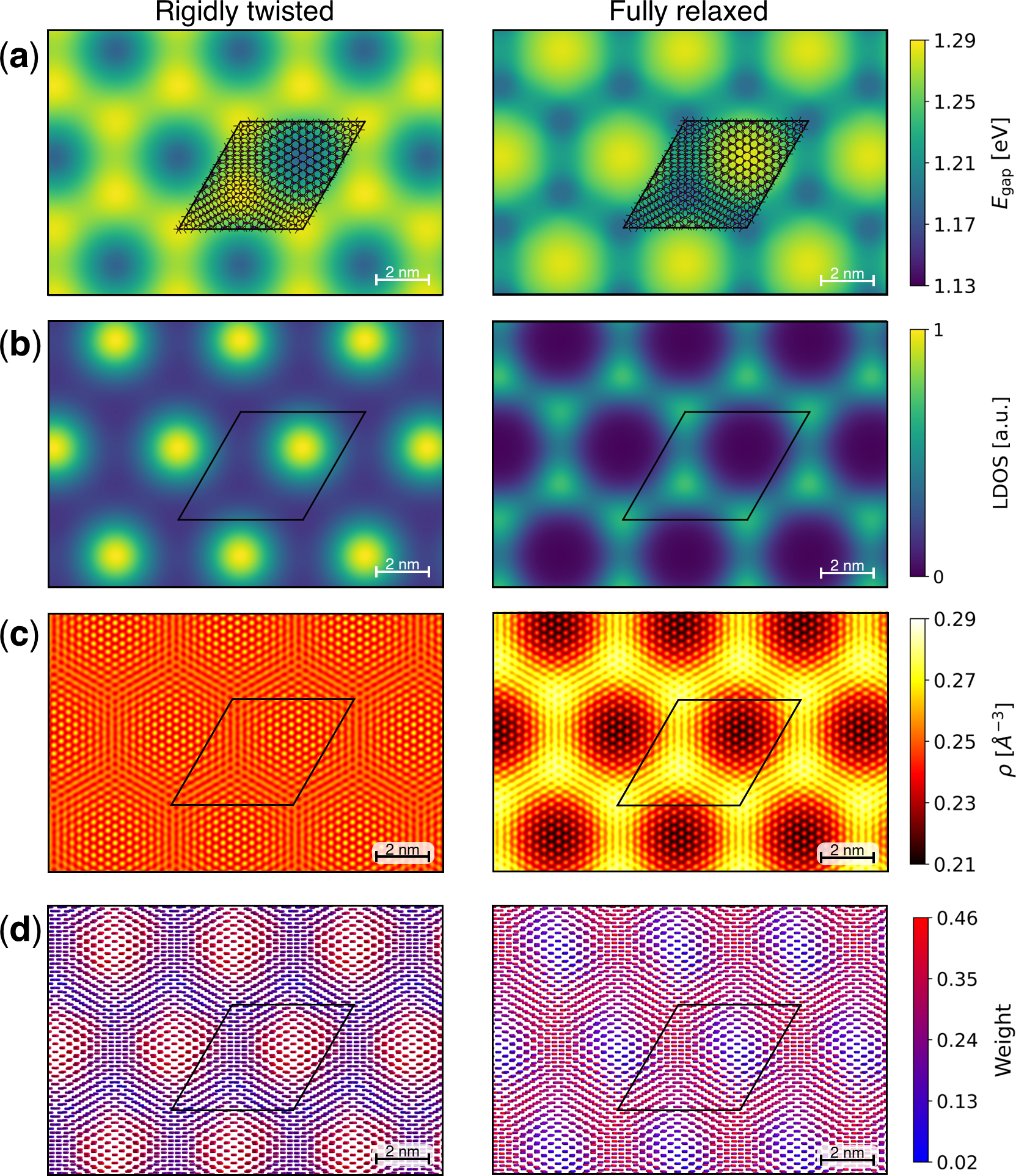}
    \caption{Local electronic properties in (left) rigidly twisted and (right) fully relaxed structures: (a) local gap, (b) atom-projected DOS (LDOS) at the VBM, (c) electron density, $\rho$, at the VBM, and (d) transmission pathways in tBL MoS\textsubscript{2} with $\theta=3.89^\circ$. The moiré cell is indicated. Plots (a) and (b) are interpolated between the data points. The atomic structure is added to subplot (a) to show the placement of structural elements. The results shown in (a) and (b) are calculated at the DFT PBE level, while (c) and (d) at the DFTB level.}
    \label{figure_local_electronic_structure}
\end{figure*}

These results show that the atomic reconstruction results in changes in the electronic properties corresponding to lattice types not present in the initial atomic structure.
To better understand the origin of these changes and their consequences in the soliton regimes, we analyzed the local electronic properties of an exemplary structure.
We used $\theta=3.89^\circ$ as the smallest twist angle which is reliably accessible both in DFTB and DFT.
The properties studied are:
(i) The local gap, a property similar to the transport gap, as shown in Fig.~\ref{figure_local_electronic_structure}(a).
It can further be linked to the moiré potential as it is affected mainly by the energy of the VBM in different parts of the structure.
Additional information is given in the SI (see Fig.~S23 and Tables~S6 and S7).
(ii) The LDOS at the VBM, as shown in Fig.~\ref{figure_local_electronic_structure}(b), with additional information on changes with $\theta$ given in the SI (see Table~S8).
Changes in this property are connected to the variation of the local gap within the structure, influencing the localization of charge carriers.
(iii) The electron density at the VBM, calculated only at the DFTB level, as shown in Fig.~\ref{figure_local_electronic_structure}(c).
This property reveals the most probable region where charge carriers are located.
(iv) The transport pathways, as shown in Fig.~\ref{figure_local_electronic_structure}(d).
It illustrates the most viable path along which charge carriers can be transported within the device configuration at the energy of the VBM.

Studying the rigidly twisted systems (left part of Fig.~\ref{figure_local_electronic_structure}) demonstrates the electronic structure obtained when structural relaxation is neglected.
The smallest value of the local gap is found in the S@S areas, while the largest ones are found in the S@Mo areas and areas of intermediate stacking.
Consequently, a hexagonal pattern of areas with increased LDOS and electron density at the VBM is formed.
The main contributions are localized in the S@S areas, corresponding to areas with the smallest local gap.
Lower contributions are found in the areas with high local gap.
This has the consequence that rigidly twisted layers would transport mainly through the S@S areas, resulting in a rather low transmission through the device.
The localized states at the VBM are reflected in the flat VB, as shown in Fig.~\ref{fig_DFTB_03.89deg}.

The observed patterns change completely when the system is allowed to relax.
The corrugated system shows the nodes as areas of the largest local gap, while the stacking domains have the smallest and the solitons intermediate values.
The range of the local gap in tBL MoS\textsubscript{2} with $\theta=3.89^\circ$ is reduced compared to the high-symmetry BL stackings with deviations of $\approx$50\,meV from both $R_h^M$ and $R_h^h$ reference systems. 
Analogous to the rigidly twisted system, areas of high (low) local gap correspond to a low (high) LDOS and reduced (increased) electron density at the VBM.
Due to this, the stacking domains and partially the solitons show increased contributions to the VBM, while the nodes do not contribute.
The fully relaxed structure in the soliton regime, thus, exhibits a honeycomb-shaped network of areas with high DOS and electron density at the VBM, corresponding to a delocalization of the holes.
This is also reported elsewhere\cite{naik2018ultraflatbands,naik2020origin,zhang2020tuning,venkateswarlu2020electronic,zhan2020tunability,vitale2021tBLtmdc} and confirmed in experiments.\cite{zhang2020flat}
Towards the transition and moiré regime, the modulation of the LDOS contributions vanishes.
A pronounced transmission pathway through the device can be observed.
This is in good agreement with the conductance calculations (see Fig.~\ref{fig_DFTB_03.89deg}), which show an increased transport in the relaxed compared to the rigidly twisted system.
The most promising direction for charge carriers to travel are the S@Mo areas and the solitons with a constant, high transmission coefficient.
The observed change in the local electronic properties throughout the structure can further be used to gain insights into the moiré potential.
As this term is usually not well-defined, the energy of the VBM (see Tables~S6 and S7) can be used as an approximation.
The stacking domains have the highest moiré potential, while it is lowered inside the nodes. 

The observed honeycomb lattice of high LDOS at the VBM can be directly linked to the observed effect of the highest two VBs forming graphene $p_z$-like bands with a Dirac cone.
The analysis of the LDOS contributions of the different structural elements at the frontier bands (see Fig.~S23(b)) supports this.
The two Dirac bands show majority contributions from the stacking domains, forming a honeycomb lattice.
The bands below show the majority contributions originating from the solitons.
This links the emergence of the kagome-type bands to a kagome lattice formed by the center of the solitons (see Fig.~\ref{fig_superlattice_bands}).
With this, the clear link between atomic reconstruction, the local electronic structure, and the emergent features in the band structure at small $\theta$ is highlighted.

\section{Conclusions}
In this work, we studied the structural, energetic, and electronic properties of tBL MoS\textsubscript{2} by changing the twist angle $\theta$ quasi-continuously.
Using a large number of fully relaxed structures and many different properties, we derive a new, comprehensive characterization of how the structure can be described in different twist angle regimes.
We further identified the $\theta$ values for which the transition between the regimes occurs.
For small twist angles $3^\circ<\theta\leq6^\circ$ ($54^\circ\leq\theta<57^\circ$) in the soliton regime, there is a strong influence of relaxation effects and an extensive atomic reconstruction, leading to a corrugation of the individual layers.
For the smallest twist angles $\theta\leq3^\circ$ ($\theta\geq57^\circ$), we describe the domain-soliton regime, where large, reconstructed stacking domains with a network of sharp domain boundaries (solitons) are found.
For large twist angles between $\approx13^\circ$ and $\approx47^\circ$ inside the moiré regime, the model of rigidly twisted layers is applicable.
Between these regimes for $6^\circ<\theta<13^\circ$ ($47^\circ<\theta<54^\circ$), the structure changes continuously, which is described by the transition regime.

We also show how the comprehensive knowledge of the structural regimes and their structural elements can be directly linked to the electronic properties and their change with $\theta$.
In rigidly twisted systems, flat bands form at small $\theta$.
When the system is allowed to relax, graphene $p_z$-like bands with Dirac cones at $K$ are found.
This originates from the emergence of extended stacking domains, forming a honeycomb superlattice of regions with low-energy stacking.
We further observe that these bands become flat for $\theta\rightarrow 0^\circ$ and that lower valence bands resemble those of a kagome lattice originating from the solitons.
These differences between the rigidly twisted and the relaxed system are also visible in the ballistic transport properties.
Analysis of the local electronic properties using the atom-projected DOS reveals a variation of the local gap of $\approx$100\,meV throughout the relaxed structure.
In the corrugated system, this originates from the states at the VBM being predominantly localized in the areas of stacking domains and solitons.
This also has consequences on the electron density and the transmission pathways.

We demonstrate that including relaxation effects is crucial when calculating the electronic structure of tBL MoS\textsubscript{2} systems.
Otherwise, the predicted properties might be wrong not only quantitatively but also qualitatively.
The model of rigidly twisted layers is only valid inside the moiré regime.
We further show that force field-based calculations allow for a computationally cheap approximation of the relaxed geometry, avoiding the computationally expensive step of geometry optimization with, e.g., DFT.
The force-field relaxed structures are, thus, reliable to use for electronic structure calculations in DFT.
We expect the findings presented here to also be important for other tBL systems beyond the studied MoS\textsubscript{2} systems.

\section{acknowledgments}
This project has been funded by Deutsche Forschungsgemeinschaft (DFG) within the Priority Program SPP 2244 ''2DMP'' and CRC 1415, and the European Union's Horizon 2020 research and innovation program under agreement No 956813.
Computational resources were provided by ZIH Dresden (project \texttt{nano-10}) and Forschungszentrum Jülich (project \texttt{opt2d}).
The authors would like to thank Roman Kempt for his technical support with the calculation of electronic properties and Louis Stuber for IT support at Forschungszentrum Jülich.

%==============%
% Bibliography %
%==============%

%apsrev4-2.bst 2019-01-14 (MD) hand-edited version of apsrev4-1.bst
%Control: key (0)
%Control: author (72) initials jnrlst
%Control: editor formatted (1) identically to author
%Control: production of article title (-1) disabled
%Control: page (0) single
%Control: year (1) truncated
%Control: production of eprint (0) enabled
%

\newpage
\onecolumngrid


\section*{Supplementary material}

# CONTENTS



**DATA AVAILABILITY**

The research data associated with this work is available as a ZENODO repository, URL: doi.org/10.5281/zenodo.7243735, [1] and contains:

- Initial, rigidly twisted geometries and fully ReaxFF-optimized geometries of 1219 tBL MoS$_2$ structures with twist angle ($\theta$) in the range of $0.2° - 59.6°$.

- Movie files, showing the change in the interlayer distance landscape and the strain fields with $\theta$.

- The Python script used to create the plots of the interlayer distance, which was also the base for plotting the local electronic properties.

- The inputs and outputs of the geometry optimization calculations with ReaxFF, using LAMMPS.

- The inputs and outputs of the geometry optimization calculations at the DFT PBE level, using FHI-aims, to verify the results from ReaxFF.

- The inputs and outputs of the electronic structure calculations at the DFT PBE level, using FHI-aims, and at the DFTB level with QN13 parameters,[2] using QuantumATK.

**INTRODUCTION**

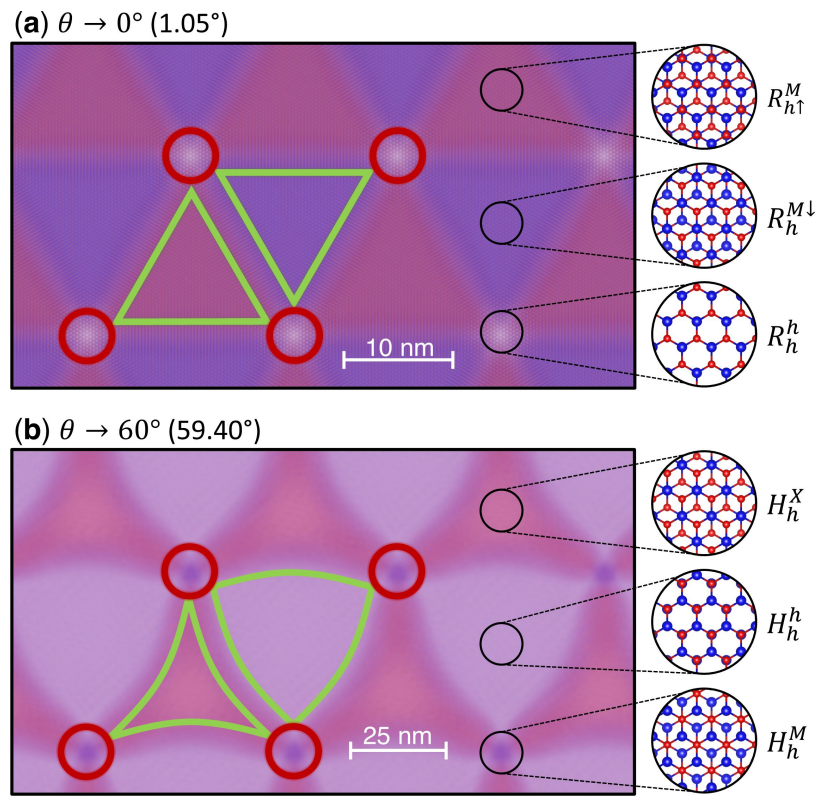

FIG. S1. Fully relaxed atomistic structure of tBL MoS$_2$ for $\theta$ close to (a) 0° and (b) 60°. The stacking domain outlines are indicated in green, and the nodes by red circles. The local BL stacking inside the structural elements is given on the right side: $R$-type for $\theta \to 0°$, $H$-type for $\theta \to 60°$

5

## METHODS

### Structure generation

Rigidly twisted BL MoS$_2$ structures without inherent interlayer strain were obtained from ideal MoS$_2$ monolayer (ML) supercells using a set of indices $m, n, p, q$ so that

$$\vec{A} = m\vec{a}_1 + n\vec{a}_2 \text{ with } m, n \in \mathbb{N}_0 \tag{1}$$

$$\vec{B} = p\vec{b}_1 + q\vec{b}_2 \text{ with } p, q \in \mathbb{N}_0 \tag{2}$$

fulfill the condition $\vec{A} = \vec{B}$ and, thus, $m^2 + mn + n^2 = p^2 + pq + q^2$. The lattice vectors $\vec{a}_i$ and $\vec{b}_i$ are of the primitive cell of ML A and B. The interlayer twist angle, $\theta$, is then given as:

$$\cos\theta = \frac{a_{\text{prim}}^2}{A^2} \left( mp + \frac{1}{2} \cdot (mq + np) + nq \right), \tag{3}$$

and the number of atoms $Z$ as:

$$Z = Z_{\text{ML}} \cdot \left( m^2 + mn + n^2 + p^2 + pq + q^2 \right) \tag{4}$$

where $Z_{\text{ML}}$ is the number of atoms in the primitive cell of the ML. Suitable sets of indices were scanned with $m + n \leq 350$ and then filtered out: For $1° \leq \theta \leq 59°$ structures were chosen with a step size of $0.01°$ and $Z \leq 100,000$. For $\theta < 1°$ and $\theta > 59°$ structures were chosen with a step size of $0.001°$ and $Z \leq 500,000$. The results are shown in Fig. S2.

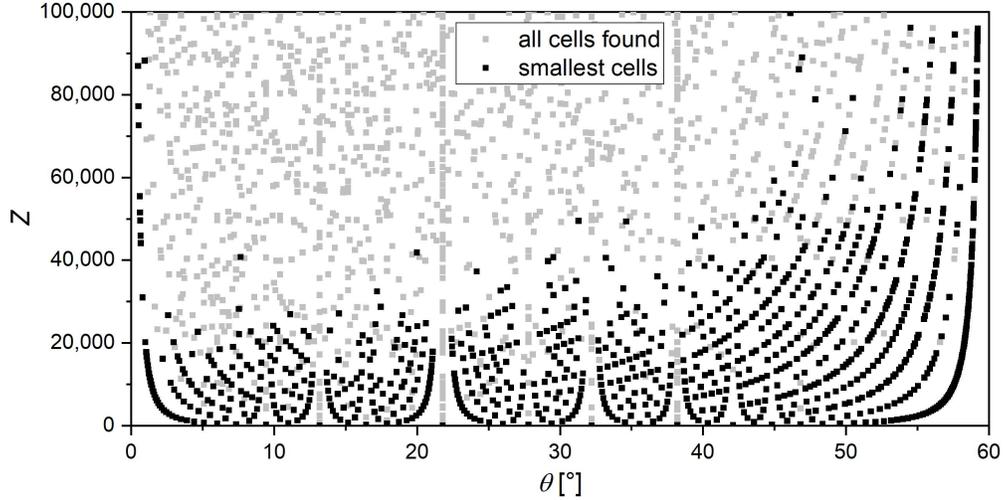

FIG. S2. Number of atoms $Z$ as a function of $\theta$ in tBL MoS$_2$ structures with no inherent interlayer strain. Grey data points represent all structures for which indices were found. Black data points are those that were used in this work after filtering them out according to the twist angle intervals.

6

It is interesting to note that the size of the simulation cell and the moiré cell only agree for a few structures. The number of moiré cells $N$ in the simulation cell is mostly larger than one. Due to the approach used to construct the initial structures from the indices $m, n, p, q$, we find that for $\theta < 30°$ the value of $N$ is given as:

$$\sqrt{N} = |m - n|.$$

(5)

The different branches in Fig. S2 originate from this effect, with each branch belonging to tBL structures with a given value of $N$. The reason is that the size of the moiré cell is determined by the continuous change in the atomic registry and, as a consequence, gives the smallest periodic unit of the local interlayer separation landscape. The simulation cell, however, also needs to fulfill periodic boundary conditions in its atomistic structure, making it necessary, in most cases, to use a moiré supercell to obtain a valid tBL structure.

**Geometry optimization with ReaxFF**

Geometry optimization was performed for all structures using ReaxFF as parameterized by Ostadhossein et al. [3] (fitted against PBE+D2) using LAMMPS [4] (version Dec. 12, 2019). Atomic coordinates were relaxed using conjugate gradients with $10^{-18}$ relative energy tolerance and a total force of $10^{-6}$ kcal/mol·Å. Cell constants of tBL $MoS_2$ were optimized by varying the cell constant $a_{\text{prim}}$ of the primitive $MoS_2$ ML around its equilibrium value of $3.186$ Å. The atomic coordinates were optimized for each initial cell constant with a fixed cell. Afterwards, a quadratic function was fitted against the total energy and a linear function against the remaining pressure. The equilibrium cell constant of the tBL $MoS_2$ cell was then calculated as the average between the minimum of the fitted total energy function and the root of the fitted pressure function. The final relaxed structure was obtained by optimizing the atomic coordinates of tBL $MoS_2$, constructed with their respective equilibrium cell constants. The geometry optimization for a given structure was considered successful if the maximum force was below $0.1$ eV/Å ($2.3$ kcal/mol·Å) and the remaining pressure did not exceed $10$ kbar ($9869$ atm). Using this procedure, a total of $1219$ fully optimized tBL $MoS_2$ structures were available for further analysis.

**Verification of ReaxFF structure with DFT**

A test set was optimized with DFT to verify that the geometry optimization with ReaxFF gives reliable results. It consists of ML $MoS_2$, the five high-symmetry BL stackings, and small tBL structures with up to 500 atoms. Calculations were carried out in FHI-vibes using the Perdew–Burke–Ernzerhof (PBE) exchange-correlation functional within the generalized gradient approximation.[5] An intermediate basis set of tier 1 was employed with a maximum force of $0.02\,\mathrm{eV/\mathring{A}}$. The lattice was allowed to relax in-plane (ip) with a fixed value of $c = 100\,\mathring{A}$. Non-local many-body dispersion (MBD) and the Tkatchenko–Scheffler (TS) method were used to treat the London dispersion interactions.[6, 7]

*Monolayer $MoS_2$*

TABLE S1. Structural properties of ML $MoS_2$, obtained from full relaxation using ReaxFF and DFT/PBE. The given values are the cell constant $a_{\mathrm{prim}}$, the Mo−S bond length, $l$, and the distance in the z-direction between two S atoms within the same layer, $\Delta z(\mathrm{S}-\mathrm{S})$. The cell constant from PBE calculations is $\approx 1\%$ smaller and the Mo-S bond is $0.03\,\mathring{A}$ shorter than in ReaxFF.

| Method | $a_{\mathrm{prim}}$ [$\mathring{A}$] | $l(\mathrm{Mo}-\mathrm{S})$ [$\mathring{A}$] | $\Delta z(\mathrm{S}-\mathrm{S})$ [$\mathring{A}$] |
|---|---|---|---|
| ReaxFF | 3.186 | 2.4324 | 3.1832 |
| PBE | 3.1528 | 2.4043 | 3.1416 |

*Bilayer $MoS_2$*

The structural properties of the high-symmetry BL $MoS_2$ stackings agree well between ReaxFF, PBE+MBD, and PBE+TS. The difference in the cell constant, $a_{\mathrm{prim}}$, and the interlayer distance, $d$, between the S@Mo and the S@S stacking types is consistent for all methods. The values of $d$ match well between ReaxFF and PBE+TS, while PBE+MBD gives values larger by $\approx 0.3\,\mathring{A}$. This is a consequence of underbinding in non-local MBD of 10% to 20% in TMDCs.[6] This difference must be considered when comparing the interlayer distances of the tBL $MoS_2$ structures between different methods, as it determines the range of possible local interlayer distances after relaxation.

TABLE S2. Structural properties of the high-symmetry $MoS_2$ BL stacking types, obtained from relaxation using ReaxFF, PBE+TS, and PBE+MBD. The given values are the cell constant, $a_{prim}$, the interlayer (Mo−Mo) distance, $d$, the Mo−S and S−S bond lengths, $l$, and the distance in the z-direction between two S atoms within the same layer, $\Delta z(S-S)$. The average compressive strain relative to the ReaxFF value is 0.75% for PBE+TS and 1% for PBE+MBD.

| Property | Method | S@Mo | | S@S | | Mo@Mo |
| | | $H_h^h$ | $R_h^M$ | $H_h^M$ | $R_h^h$ | $H_h^X$ |
|---|---|---|---|---|---|---|
| $a_{prim}$ [Å] | ReaxFF | 3.1852 | 3.1852 | 3.1844 | 3.1844 | 3.1852 |
| | PBE+TS | 3.1605 | 3.1619 | 3.1597 | 3.1595 | 3.1641 |
| | PBE+MBD | 3.1533 | 3.1542 | 3.1528 | 3.1528 | 3.1546 |
| $d$ [Å] | ReaxFF | 6.0299 | 6.0317 | 6.4985 | 6.4972 | 6.0346 |
| | PBE+TS | 6.0283 | 6.0204 | 6.6023 | 6.6423 | 6.0262 |
| | PBE+MBD | 6.2796 | 6.2693 | 6.8706 | 6.9039 | 6.3791 |
| $l(Mo-S)$ [Å] | ReaxFF | 2.4310 | 2.4310 | 2.4305 | 2.4305 | 2.4311 |
| | PBE+TS | 2.4055 | 2.4059 | 2.4062 | 2.4062 | 2.4059 |
| | PBE+MBD | 2.4034 | 2.4037 | 2.4034 | 2.4034 | 2.4038 |
| $l(S-S)$ [Å] | ReaxFF | 2.8513 | 2.8530 | 3.3200 | 3.3188 | 2.8558 |
| | PBE+TS | 2.8988 | 2.8932 | 3.4680 | 3.5076 | 2.9043 |
| | PBE+MBD | 3.1414 | 3.1322 | 3.7315 | 3.7648 | 3.2428 |
| $\Delta z(S-S)$ [Å] | ReaxFF | 3.1800 | 3.1800 | 3.1795 | 3.1794 | 3.1801 |
| | PBE+TS | 3.1350 | 3.1342 | 3.1382 | 3.1385 | 3.1312 |
| | PBE+MBD | 3.1382 | 3.1379 | 3.1389 | 3.1389 | 3.1377 |

*Twisted bilayer $MoS_2$*

The relative cell constant (see Fig. S3(a)) shows a consistent behavior between ReaxFF and PBE+vdW with an exception at 60° ($H_h^h$ BL). The mean interlayer distance (see Fig. S3(c)) shows an excellent agreement between ReaxFF and PBE+TS. PBE+MBD gives a larger $d$, analogous to the BL stackings. However, $d_{mean}$ relative to the 0° low-energy stacking type $R_h^M$ (see Fig. S3(d)) shows a very good agreement between all methods. The main difference is the curvature of the individual lines, indicating differences in the deforma-

tion behavior of the layers. This is approximated by the variation of the $z$-coordinate of the Mo atoms inside the layers, $\Delta z$ (see Fig. S3(b)). ReaxFF shows the largest $\Delta z$ and, thus, the strongest waviness of the layers. This can be understood by the layers being described as less rigid in ReaxFF compared to PBE+vdW. Consequently, the layers start to deform at larger $\theta$ and, thus, the $\theta$ values for the transition between the different regimes can be expected to vary.

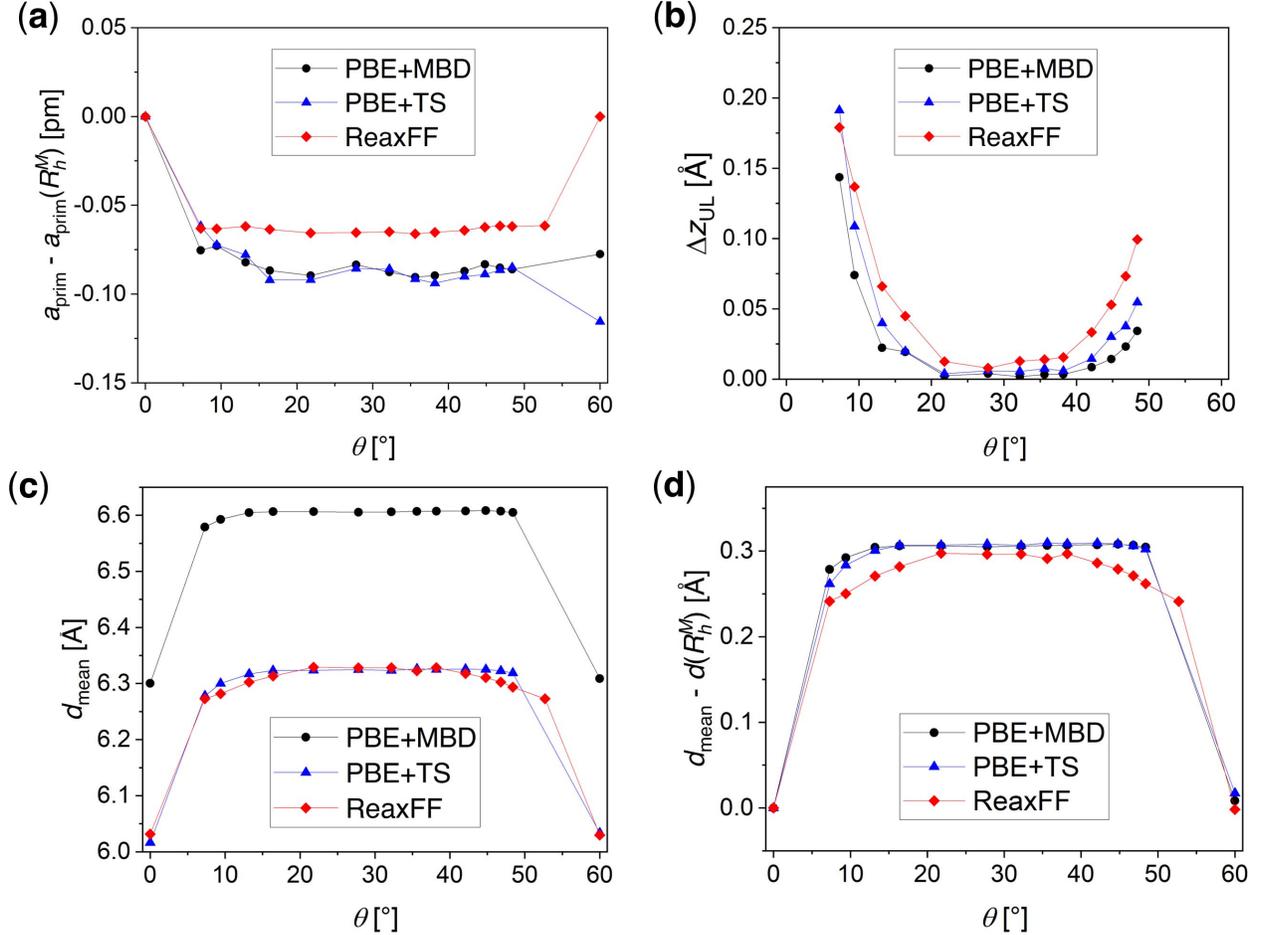

FIG. S3. Structural properties of tBL $MoS_2$ as a function of $\theta$ for structures relaxed with different methods: (a) Primitive cell constant, $a_{prim}$, relative to the value of the $R_h^M$ stacking. (b) Difference between the minimum and maximum $z$-coordinate, $\Delta z_{UL}$, of the Mo atoms inside the upper layer (UL). $\Delta z_{LL}$ inside the lower layer (LL) gives almost identical results. (c) Mean interlayer distance, $d_{mean}$. (d) Mean interlayer distance, $d_{mean}$, relative to the value of the $R_h^M$ stacking.

**Analysis of structural properties**

Our analysis is based on the moiré cell, which we define as the smallest repeating unit in the local interlayer separation landscape of the simulation cell. We construct it from the positions of the nodes as areas of maximum local interlayer distance, placed at the vertices. The solitons are located on the edges and the short diagonal of the moiré cell. This allows for a description independent of the original simulation cell and for an easier analysis of the structural elements. The general procedure is as follows:

1. Import structure (assumption: oriented in $x, y$-plane).

2. Identify Mo atoms and sort them into upper and lower layers.

3. Calculate the local interlayer distance via interpolation:

   - Extend Mo atoms into supercell to avoid edge effects during interpolation (extending by 20% of lattice constant or 35 Å in each direction is sufficient).
   - Interpolate $z$-coordinates of Mo atoms at points of interpolation grid $(x, y)$ with suitable step size for both layers (`interpolate.griddata` from SciPy with `method=cubic` [8] is used in this work).
   - Obtain local interlayer distance $d$ from $d(x, y) = z_{\mathrm{UL}} - z_{\mathrm{LL}}$.

4. Identify the moiré cell:

   - Find position and number $N$ of local maxima (=nodes) in local interlayer distance.
   - If $N = 1$, use the position of nodes as origin of moiré cell, and apply lattice parameters of original cell.
   - If $N > 1$, use global maximum of $d$ as origin of moiré cell (reference node), and use its nearest neighbor nodes to construct the hexagonal moiré cell.

5. Interpolate along path in the moiré cell:

   - Path $\vec{p}$ in fractional coordinates of moiré cell: $(-1/3, -1/3) \rightarrow (2/3, 2/3)$.
   - Create interpolation points $(x, y)$ along $\vec{p}$.
   - Interpolate $z$-coordinates of Mo atoms at $(x, y)$ for both layers.
   - Calculate $d$ along $\vec{p}$.

6. Analyse structure of nodes and solitons:

- Determine cross-section of nodes and solitons from interpolation along $\vec{p}$.
- Calculate height $h$ from maximum of $d$ in cross-section and global minimum of $d$ inside the simulation cell.
- Fit Gaussian $f(\vec{p}) = h \cdot \exp\left(-\frac{(p-\mu)^2}{2\sigma^2}\right)$ against cross-section along $\vec{p}$.
- Calculate base width $w$ as $4\sigma$.

The results for an interpolation along the path $\vec{p}$ to obtain the cross-section of the nodes and solitons are shown for two examples in Fig. S4. Fitting a Gaussian to approximate their shape is based on the findings of Carmesin et al.[9], who describe the curvature of TMDC BLs as being Gaussian. The consistently high $R^2$ values ($R^2 \geq 0.97$) for all structures with pronounced structural elements confirm this approach. We use the base width $4\sigma$ as a characteristic property of Gaussians to determine the spatial extension of the structural elements. This approach tends to overestimate the size of nodes and solitons slightly.

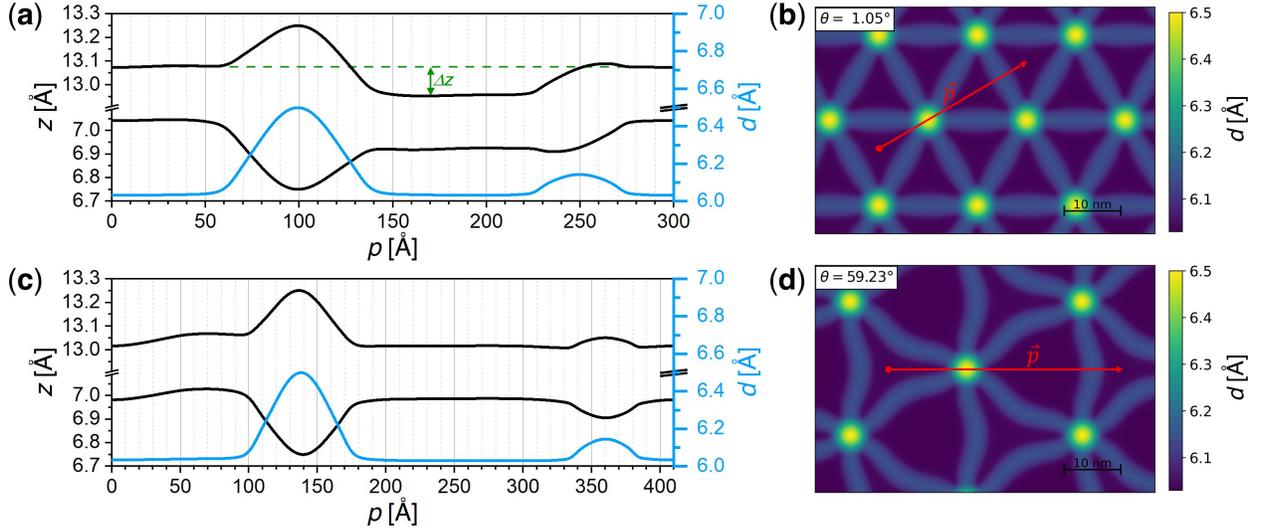

FIG. S4. (a,c) Individual layer $z$-coordinates (primary axis, with axis break) along a path $\vec{p}$ through the structural elements in the relaxed cell, including the local interlayer distance (secondary axis) for (a) $\theta = 1.05°$ and (c) $\theta = 59.23°$. (b,d) Local interlayer distance and the placement of $\vec{p}$.

We chose the interpolation method based on extensive tests of the tools for multivariate interpolation available in the SciPy library. Both different radial basis functions and the `interpolate.griddata` finite element method with cubic functions give adequate fit results during the path interpolation and similar values for the size of the structural elements. As

the deviations of the base width were below 0.01 Å, we chose the method with the shortest calculation time. We experienced no emergence of interpolation artifacts during this study.

The size of the stacking domains was approximated based on the cell constant of the hexagonal moiré cell, $a_M$, and the diameter of nodes, $d_N$, and solitons, $d_S$. It assumes a structure corresponding to tBL with $\theta \to 0°$, circular nodes, and that the solitons have a constant width along their entire length. The changes in $\theta \to 60°$ with bent solitons and differently sized stacking domains are not accounted for but instead assumed to cancel out as half of the stacking domain increases in size while the other half gets smaller. If the conditions $d_N < \frac{2}{\sqrt{3}}a_M$ and $d_S < \frac{1}{\sqrt{3}}a_M$ are fulfilled, the area of the stacking domains is non-zero and is calculated from simple geometrical considerations.

**Energy contributions**

Both binding energy, $E_{\text{binding}}$, and exfoliation energy, $E_{\text{exfol}}$, are commonly encountered terms. They differ in sign, but are both calculated from the energy of the fully relaxed tBL system, $E_{\text{tBL}}^{\text{relaxed}}$, and the free-standing, unstrained monolayers, $E_{\text{ML}}^{\text{ideal}}$, using:

$$E_{\text{binding}} = E_{\text{tBL}}^{\text{relaxed}} - \left(E_{\text{ML,A}}^{\text{ideal}} + E_{\text{ML,B}}^{\text{ideal}}\right) = -E_{\text{exfol}}. \tag{6}$$

The separation into individual energy contributions is fully mathematical and is not possible in real systems. It assumes three processes taking place consecutively (see Fig. S5).

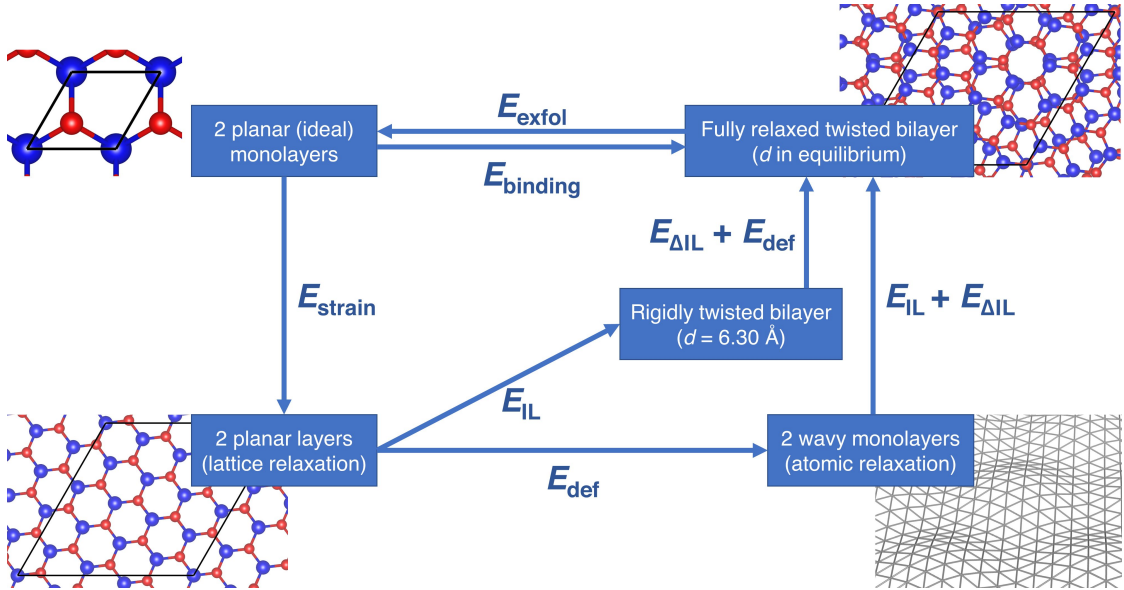

FIG. S5. Schematic representation of the energy contributions for the creation of fully relaxed tBL MoS$_2$ from freestanding, unstrained ML MoS$_2$.

Starting point are the MoS$_2$ ML supercells used to construct the initial, rigidly twisted BL systems with the cell constant of ideal, freestanding ML MoS$_2$. The cell constant of the ML is then changed to the value obtained during the lattice relaxation, introducing an intralayer strain. The strain energy $E_{\text{strain}}$ is obtained as:

$$E_{\text{strain}} = \left(E_{\text{ML,A}}^{\text{strained}} + E_{\text{ML,B}}^{\text{strained}}\right) - \left(E_{\text{ML,A}}^{\text{ideal}} + E_{\text{ML,B}}^{\text{ideal}}\right) \tag{7}$$

from the total energy of the flat ML used to construct the rigidly twisted BL with ideal and optimized (strained) cell constant. In the next step, the individual MLs, still separated, are

deformed by the relaxation of the atomic coordinates, introducing an intralayer waviness. The deformation energy, $E_{\text{def}}$, is defined as:

$$E_{\text{def}} = \left( E_{\text{ML,A}}^{\text{optim}} + E_{\text{ML,B}}^{\text{optim}} \right) - \left( E_{\text{ML,A}}^{\text{strained}} + E_{\text{ML,B}}^{\text{strained}} \right) \tag{8}$$

with $E_{\text{ML},i}^{\text{optim}}$ as the total energy of the individual layers in the fully relaxed tBL system. Finally, the monolayers are bound by the interlayer interactions to form the fully relaxed tBL system. We further separate this adhesion contribution into two parts: The first part originates from the adhesion between the flat MLs in the rigidly twisted system. This interlayer (IL) adhesion energy, $E_{\text{IL}}$, is obtained from:

$$E_{\text{IL}} = E_{\text{tBL}}^{\text{rigid}} - E_{\text{ML,A}}^{\text{strained}} - E_{\text{ML,B}}^{\text{strained}} \tag{9}$$

using the total energy of the rigidly twisted tBL system, $E_{\text{tBL}}^{\text{rigid}}$, with an interlayer distance of 6.30 Å (intermediate between the high-symmetry BL stackings). The second part is the change in the interlayer adhesion when the individual layers relax. This change in the IL adhesion energy, $E_{\Delta\text{IL}}$ is obtained from:

$$E_{\Delta\text{IL}} = E_{\text{tBL}}^{\text{relaxed}} - E_{\text{tBL}}^{\text{rigid}} - E_{\text{def}}. \tag{10}$$

To allow for a comparison between different tBL $MoS_2$ systems, the energies were normalized as follows:

$$e_i = \frac{E_i}{\sin 60^\circ \cdot a_{\text{tBL}}^2} \tag{11}$$

with $a_{\text{tBL}}$ as the simulation cell constant.

**Band structure: ReaxFF vs. DFT geometries**

*Bilayer MoS₂*

The PBE band structure of the BLs relaxed with different methods is in good agreement (see Fig. S6). The indirect gap is consistently reproduced with the valence band maximum (VBM) positioned at Γ. The conduction band minimum (CBM) position is less consistent: while it is at $Q$ for all ReaxFF structures, it is at $K$ for some PBE+vdW structures. The small energy difference of the CB at $K$ and $Q$ makes the position of the CBM sensitive to the stacking type, strain effects, and changes in the Mo−S bond length between ReaxFF and PBE-optimized structures.

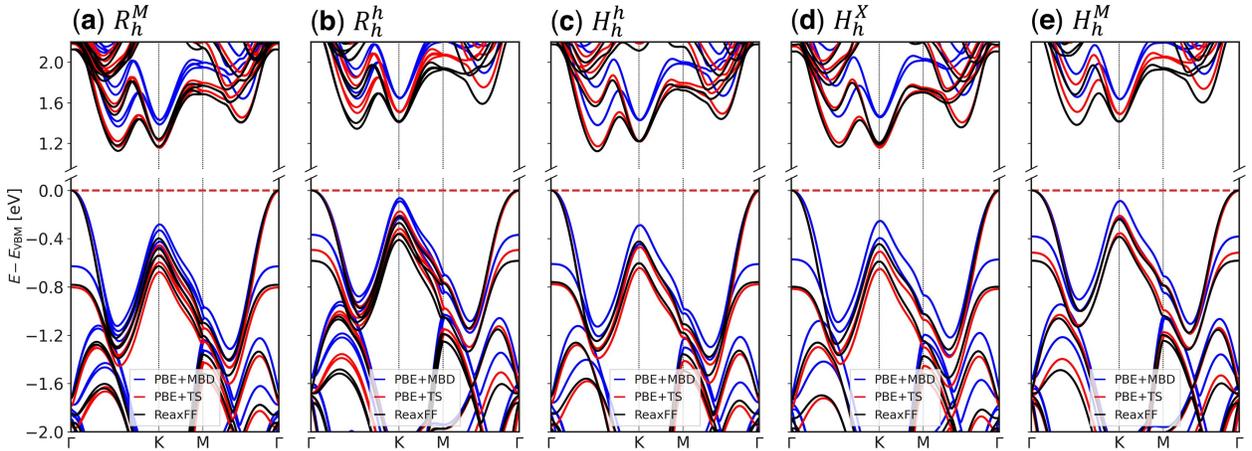

FIG. S6. Frontier bands in BL MoS₂ for the high-symmetry stackings using geometries obtained by relaxation with ReaxFF, PBE+TS, and PBE+MBD.

Except for this discrepancy, the band structures agree well between all stacking types and methods. Generally, the band structures and band gaps (see Fig. S7) calculated for the ReaxFF structures are close to those of the PBE+TS structures, while the PBE+MBD structures deviate stronger with a consistent shift towards larger band gaps. In conclusion, the results obtained for the fully ReaxFF-optimized BL MoS₂ structures are consistent with those obtained from optimization with PBE+vdW.

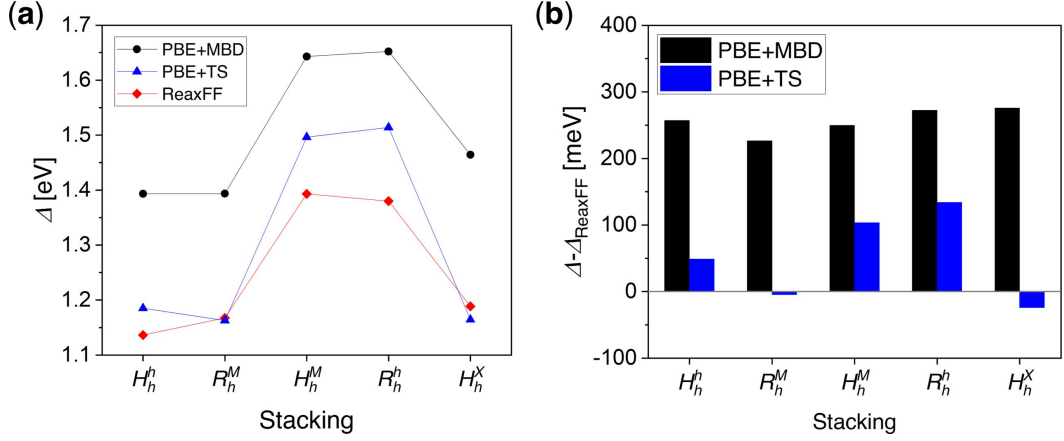

FIG. S7. Band gap of the different MoS$_2$ BL stackings. (a) Absolute values and (b) relative to those obtained for the ReaxFF-optimized structures.

*Twisted bilayer MoS$_2$*

In tBL systems, the band structures calculated using geometries optimized with ReaxFF and PBE+vdW also agree well. Identical to the BL systems, the band structures based on the PBE+TS geometry match best with the ReaxFF-based band structures. Thus, the ReaxFF-optimized geometry can be used to avoid a computationally expensive geometry optimization using DFT.

TABLE S3: Band structure of tBL MoS$_2$ with different twist angles $\theta$ for (from left to right column) the rigidly twisted structure ($d = 6.30$ Å), the ReaxFF-optimized structure, the structure optimized with PBE+TS and with PBE+MBD.

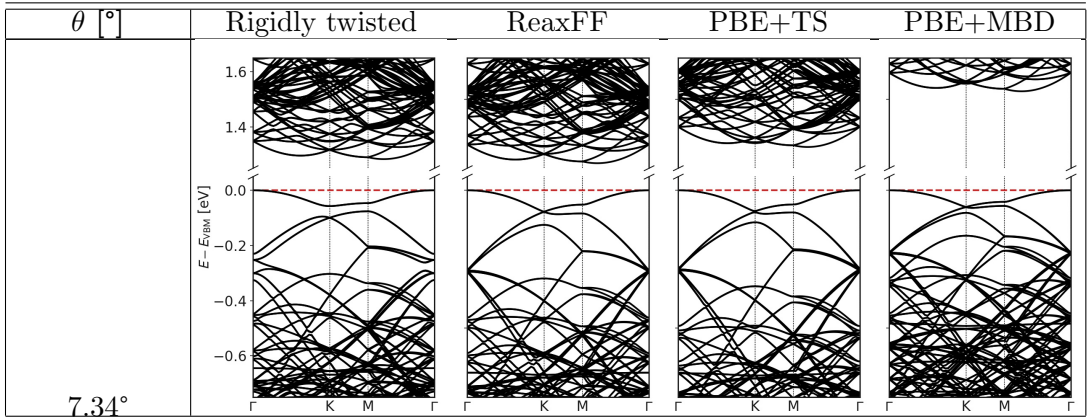

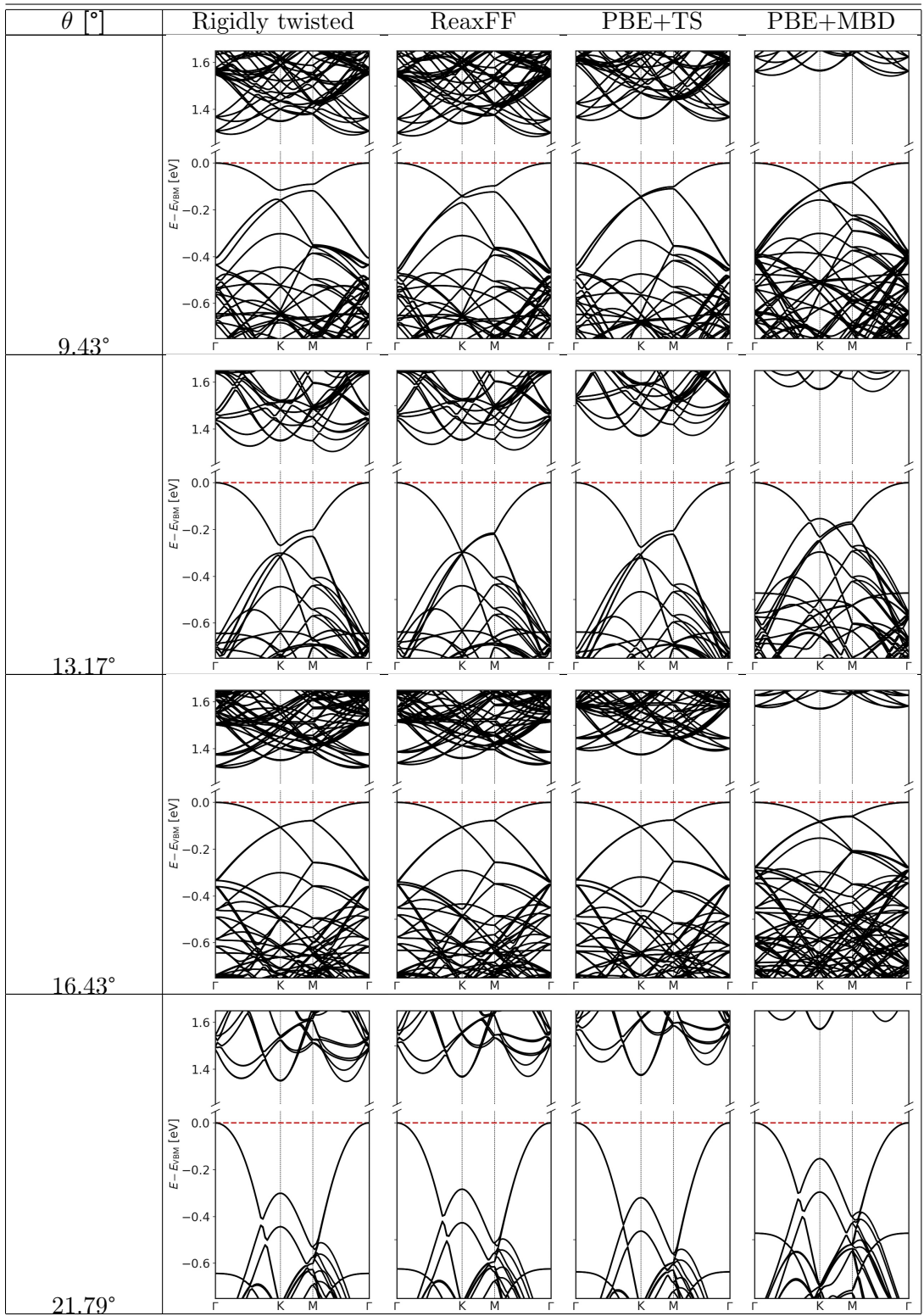

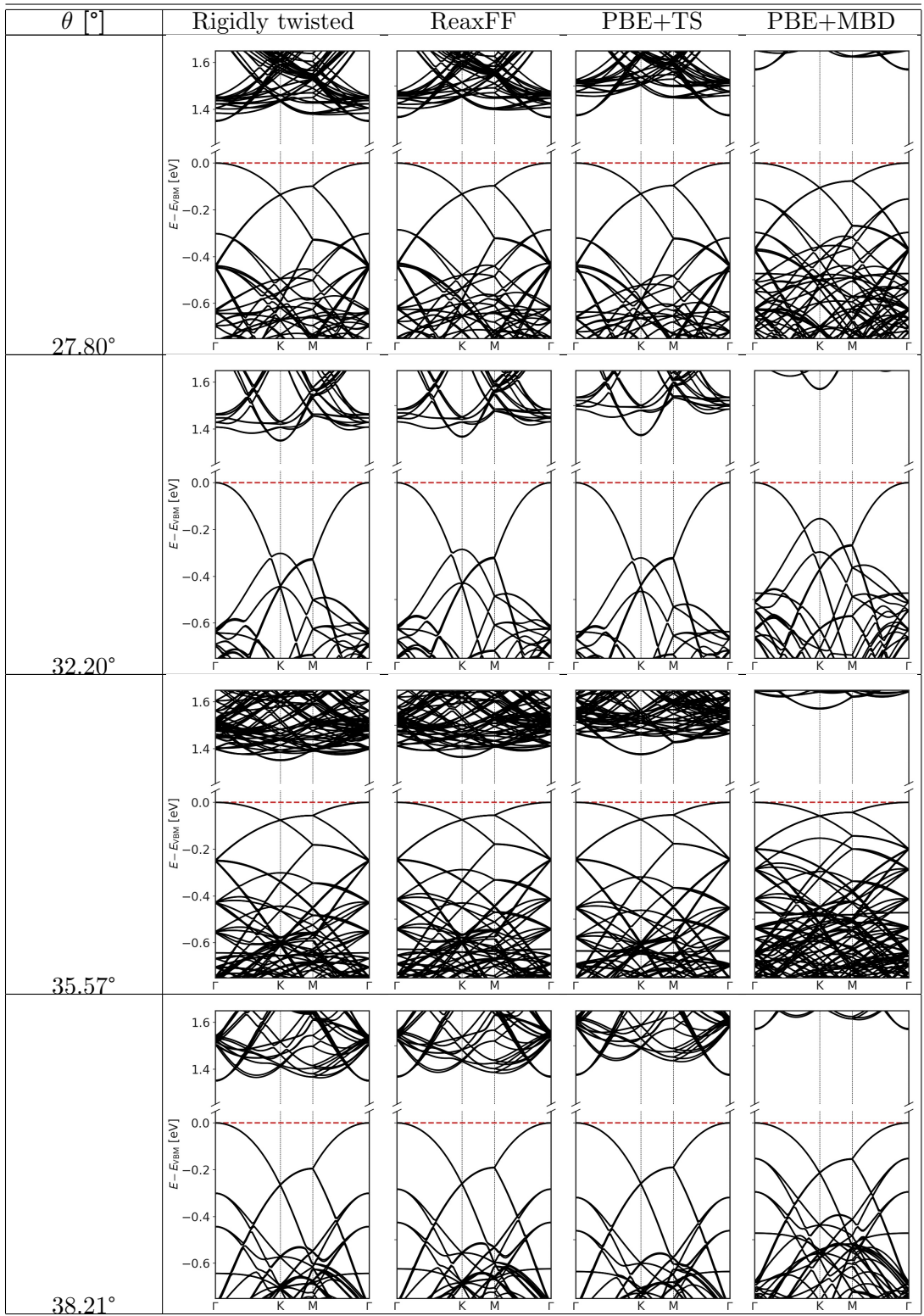

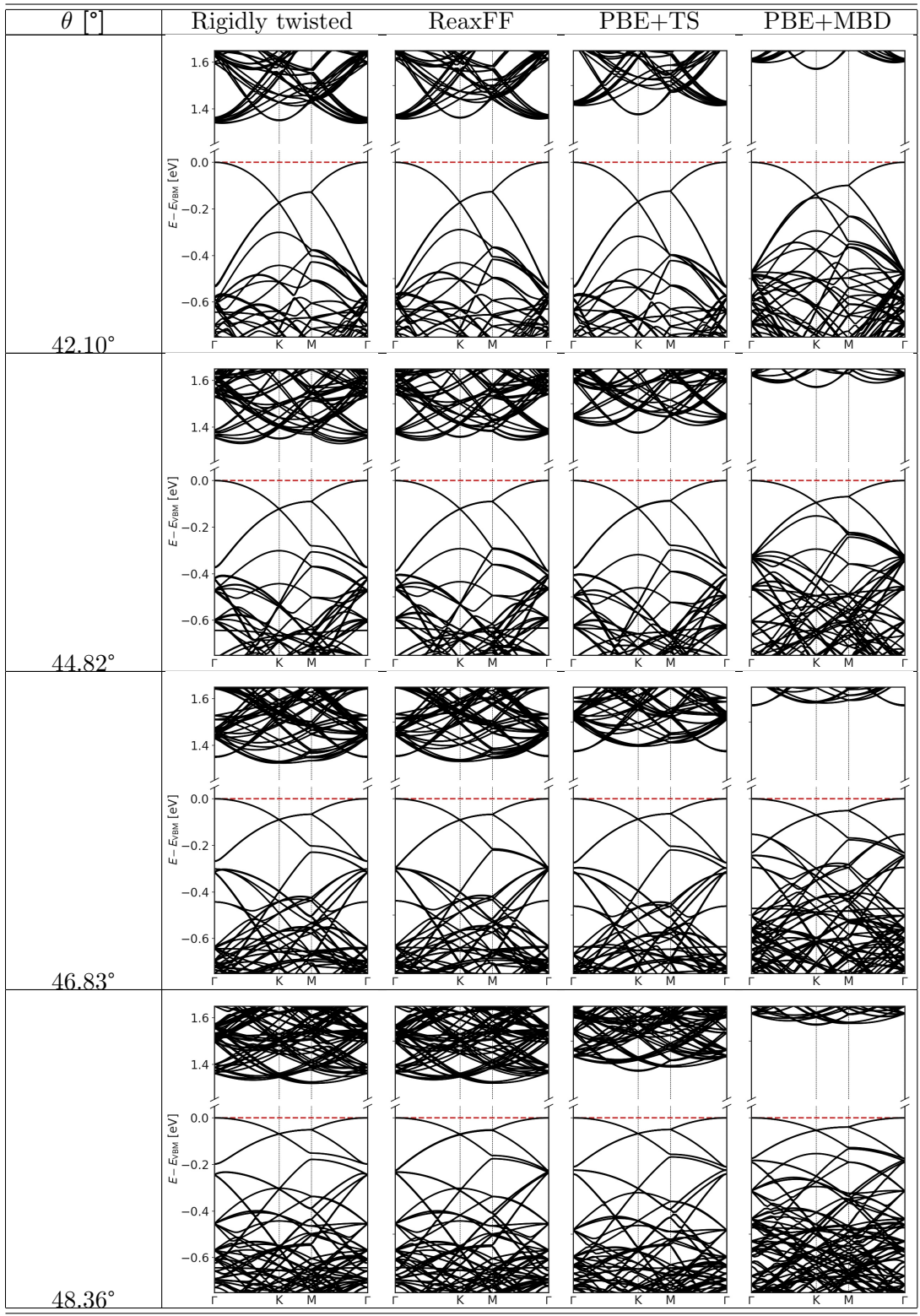

The band gaps for tBL MoS$_2$ structures obtained from different methods show a qualitative agreement (see Fig. S8). While PBE+MBD tends to give a larger absolute band gap, the values obtained for PBE+TS structures match well with those obtained for ReaxFF structures – this is similar to what is observed for BL MoS$_2$. The band gaps relative to the $R_h^M$ stacking give a very good agreement between all the methods, with a variation of up to $\approx 70$ meV. However, the area of a constant band gap is larger in PBE+vdW compared to ReaxFF structures. Furthermore, in rigidly twisted and ReaxFF-optimized structures, the decrease at small $\theta$ is almost linear, while it is nonlinear in the PBE+vdW-optimized structures. This shows that the ReaxFF geometry is usable for studying the trends with changing $\theta$ but cannot fully reproduce all details.

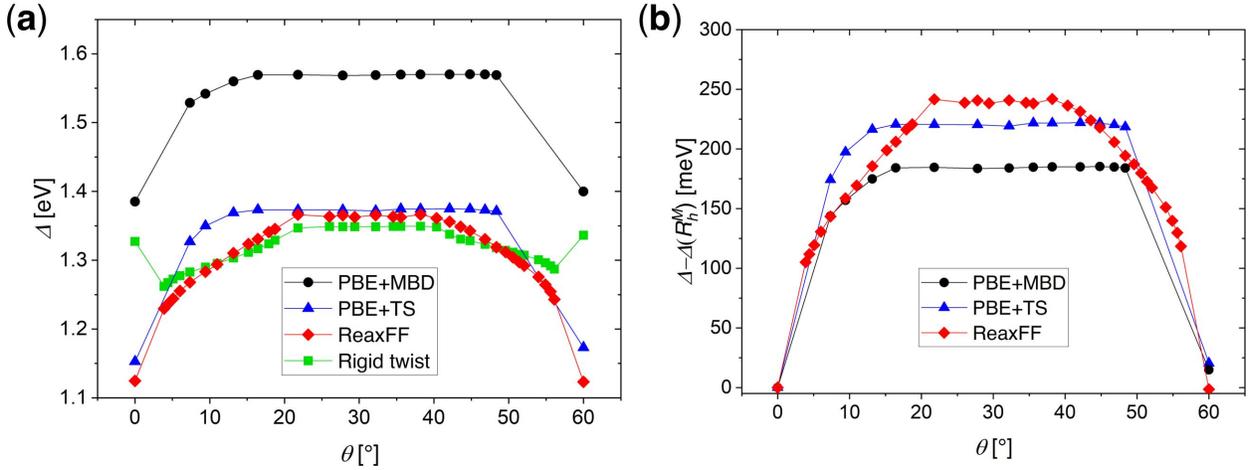

FIG. S8. Band gap $\Delta$ for tBL MoS$_2$ as a function of $\theta$ using rigidly twisted and fully relaxed (ReaxFF, PBE+TS, and PBE+MBD) structures: (a) Absolute values and (b) values relative to the value of the $R_h^M$ stacking. The rigidly twisted layers are excluded in (b) as the BL without optimization of the interlayer distance results in a too strongly deviating band gap.

**DFTB and transport calculations**

All transport calculations were performed on density-functional based tight-binding (DFTB) level. A comparison between calculations at the DFT and DFTB levels is given in Figs. S9 and S10, in addition to Fig. 7. Good agreement was obtained, supporting the choice of DFTB for transport calculations. The variation of the indirect band gap with $\theta$ is similar within the framework of DFTB and DFT, as shown in Fig. S11.

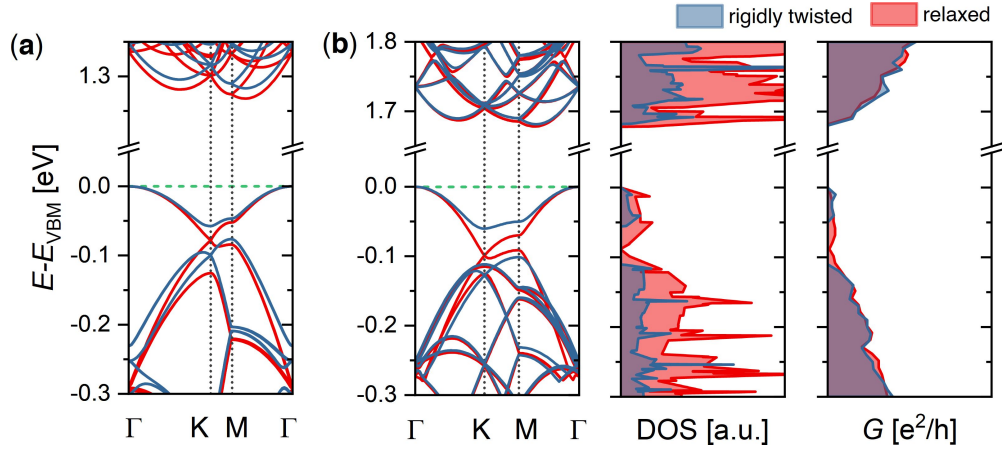

FIG. S9. (a) DFT and (b) DFTB band structure, DOS, and conductance in tBL $MoS_2$ with $\theta = 7.34°$. The results for the rigidly twisted structure are shown in blue, and for the relaxed structure in red.

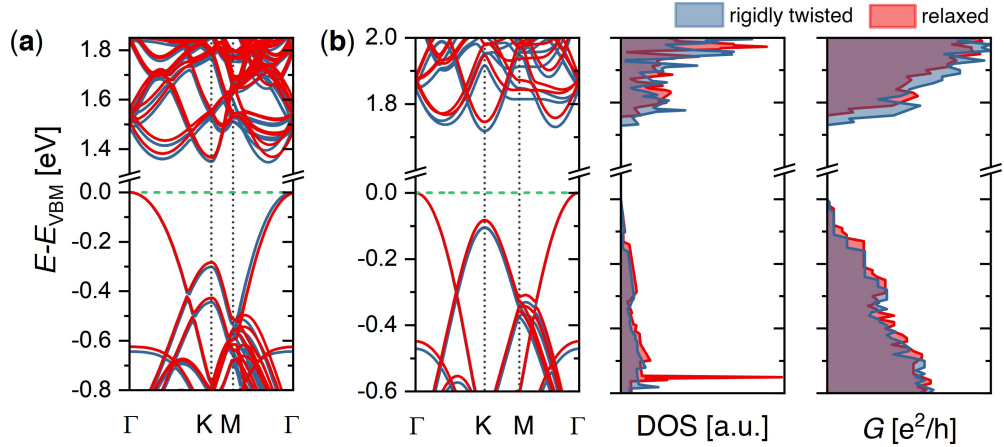

FIG. S10. (a) DFT and (b) DFTB band structure, DOS, and conductance in tBL $MoS_2$ with $\theta = 21.79°$. The results for the rigidly twisted structure are shown in blue, and for the relaxed structure in red.

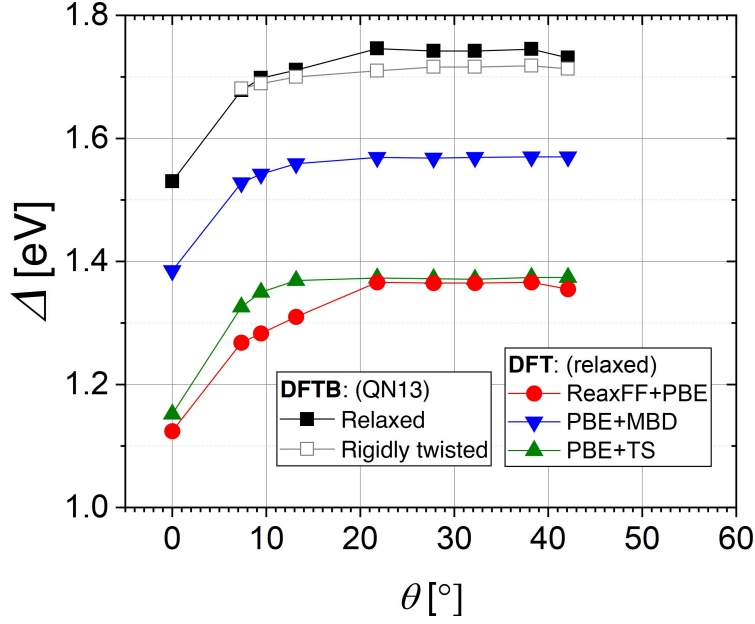

FIG. S11. Indirect band gap $\Delta$ as a function of $\theta$ at DFTB level (QN13 parameter set, rigidly twisted vs. ReaxFF-optimized geometry) compared with DFT calculations using DFT- and ReaxFF-optimized geometries.

The quantum transmission was computed at zero bias using the Landauer-Büttiker and NEGF approach [10, 11]

$$T(E) = \frac{2e^2}{h} tr[G^\dagger \Gamma_R G \Gamma_L], \tag{12}$$

where $G$ indicates the total Green's function of the scattering region coupled to the electrodes. $\Gamma_\alpha = -2Im(\Sigma_\alpha)$ is the broadening function. $\Sigma_\alpha$ with $\alpha = L, R$ are self-energies, calculated through self-consistent iteration.[12] Transmission pathways [13] were calculated to allow for an analysis option that splits the transmission coefficient into local bond contributions, $T_{i,j}$. If the system is divided into two parts (A, B), the pathways are across the boundary between A and B, summing up the total transmission coefficient

$$T(E) = \sum_{i \in A, j \in B} T_{i,j}(E). \tag{13}$$

The direction of the pathway (either from $i$ to $j$, or $j$ to $i$) shows the flow of the electrons in the device. The device configuration used in these calculations is shown for an exemplary system with $\theta = 3.89°$ in Fig. S12.

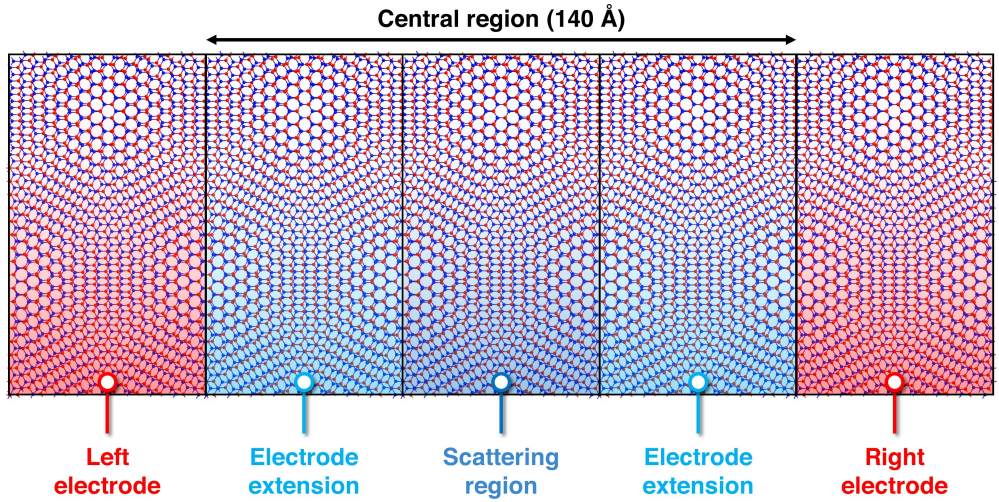

FIG. S12. Schematic representation of the device configuration for $\theta = 3.89°$ that is split into three parts. The central region of the device is finite (non-periodic) along the transport direction. It is attached to the left and right semi-infinite (periodic) electrodes, the source and drain, respectively. The transport direction is from left to right.

# STRUCTURAL AND ENERGETIC PROPERTIES

## MoS$_2$ BL stacking types

TABLE S4. Structural and energetic properties of BL MoS$_2$, fully optimized with the ReaxFF parameters by Ostadhossein et al.[3]: cell constant $a$ (ML MoS$_2$ has $a = 3.186$ Å), interlayer distance $d$, and total energy per unit cell (uc) relative to the lowest-energy stacking $\Delta E_{tot}$.

| Stacking | $H_h^h$ | $R_h^M$ | $H_h^M$ | $R_h^h$ | $H_h^X$ |
|---|---|---|---|---|---|
| Stacking type | S@Mo | | S@S | | Mo@Mo |
| $a$ [Å] | 3.1852 | 3.1852 | 3.1844 | 3.1844 | 3.1852 |
| $d$ [Å] | 6.03 | 6.03 | 6.50 | 6.50 | 6.04 |
| $\Delta E_{tot}$ [meV/uc] | 0 | 1.6 | 53.7 | 52.5 | 2.0 |

**Structure of fully-optimized tBL MoS$_2$**

*Cell constant*

The cell constant of tBL MoS$_2$ as a function of $\theta$ is studied by projecting the cell constant of the simulation cell, $A$, to the cell constant $a_{\text{prim}}$ of a primitive MoS$_2$ ML using

$$A = a_{\text{prim}}\sqrt{m^2 + mn + n^2}. \tag{14}$$

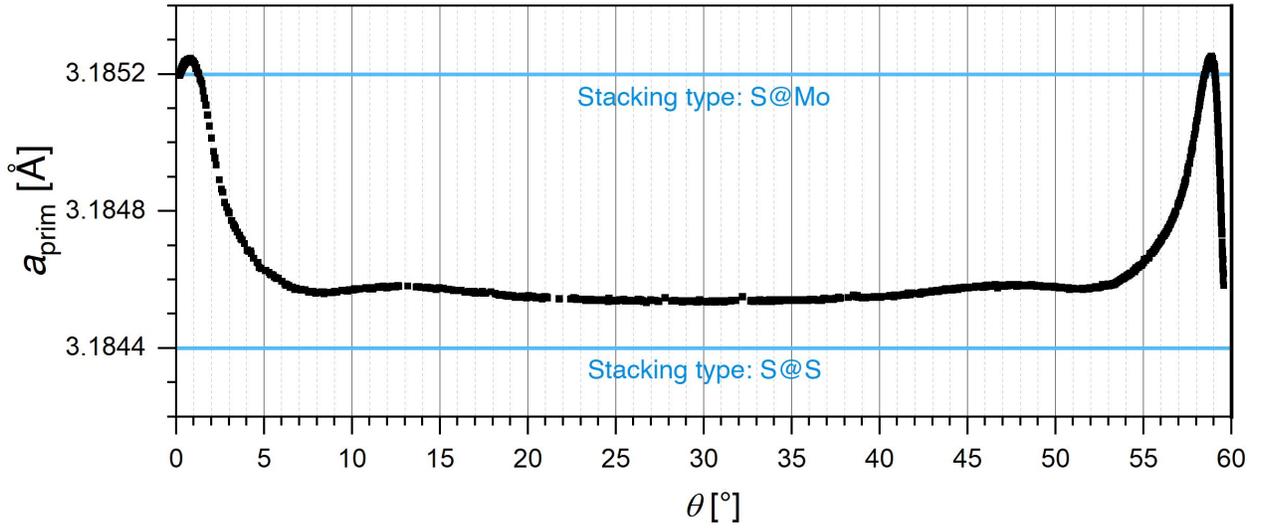

FIG. S13. Projected primitive cell constant $a_{\text{prim}}$ in fully optimized tBL MoS$_2$ as a function of $\theta$. The cell constant of the S@S and S@Mo BL stacking types is indicated for comparison. Compared to the freestanding ML ($a_{\text{ML}} = 3.186$ Å), the cell constant in tBL MoS$_2$ is compressed by a maximum of 0.05%. Even though these changes are small, we found that optimizing the cell constant was necessary to obtain sufficiently relaxed structures that satisfy the convergence conditions.

*Moiré periodicity*

The moiré periodicity $a_M$ is the cell constant of the periodic unit in the moiré pattern. It can be evaluated numerically from the local interlayer distance landscape or analytically as

$$a_M = \frac{a_{prim}}{\sqrt{2 \cdot (1 - \cos \theta')}} \text{ with } \theta' = \begin{cases} \theta & \text{for } \theta \leq 30° \\ 60° - \theta & \text{for } \theta > 30° \end{cases}. \tag{15}$$

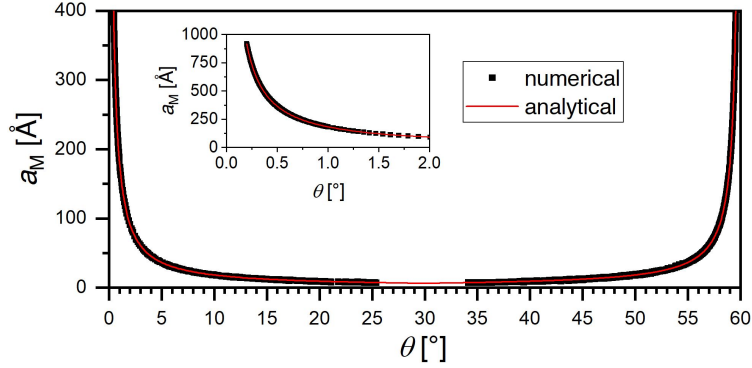

FIG. S14. Moiré periodicity $a_M$ in tBL MoS$_2$. Numerical values are determined for the fully relaxed structures; analytical values are calculated using $a_{prim} = 3.186\,\text{Å}$. This confirms that the numerical method to determine the moiré cell successfully reproduces the analytical value of $a_M$.

*Size of stacking domains*

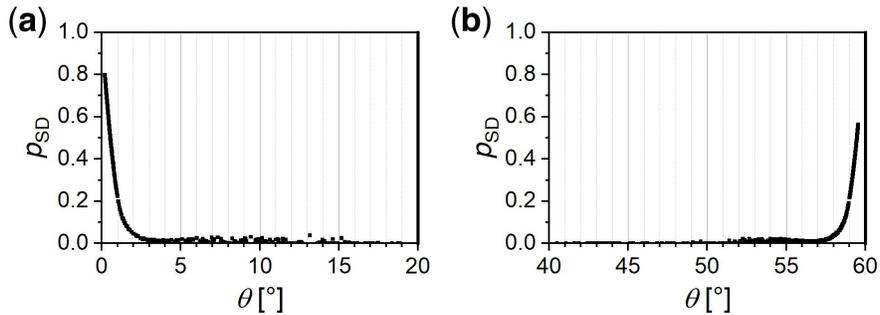

FIG. S15. Approximated fraction $p_{SD}$ of the moiré cell accounted for the stacking domains in tBL MoS$_2$ with (a) $0° < \theta \leq 20°$ and (b) $40° \leq \theta < 60°$. The geometric model can identify no stacking domains for the interval $14° < \theta < 48°$. While $p_{SD}$ strongly increases towards $0°$ ($60°$), a value of $100\%$ can only be found in the limit of an infinitely large moiré cell, where the size of nodes and solitons becomes negligibly small, effectively resembling an untwisted S@Mo BL.

## Exfoliation energy

*Energy contributions*

The strain energy follows a quadratic function of $a_{\mathrm{prim}}$ ($R^2 = 0.99986$). This confirms that the change of the cell constant is small enough to be described as harmonic and verifies the approach used for optimizing the cell constant. Both $e_{\Delta\mathrm{IL}}$ ($R^2 > 0.9999$) and $e_{\mathrm{IL}}$ ($R^2 = 0.98$) change linearly with $d_{\mathrm{mean}}$.

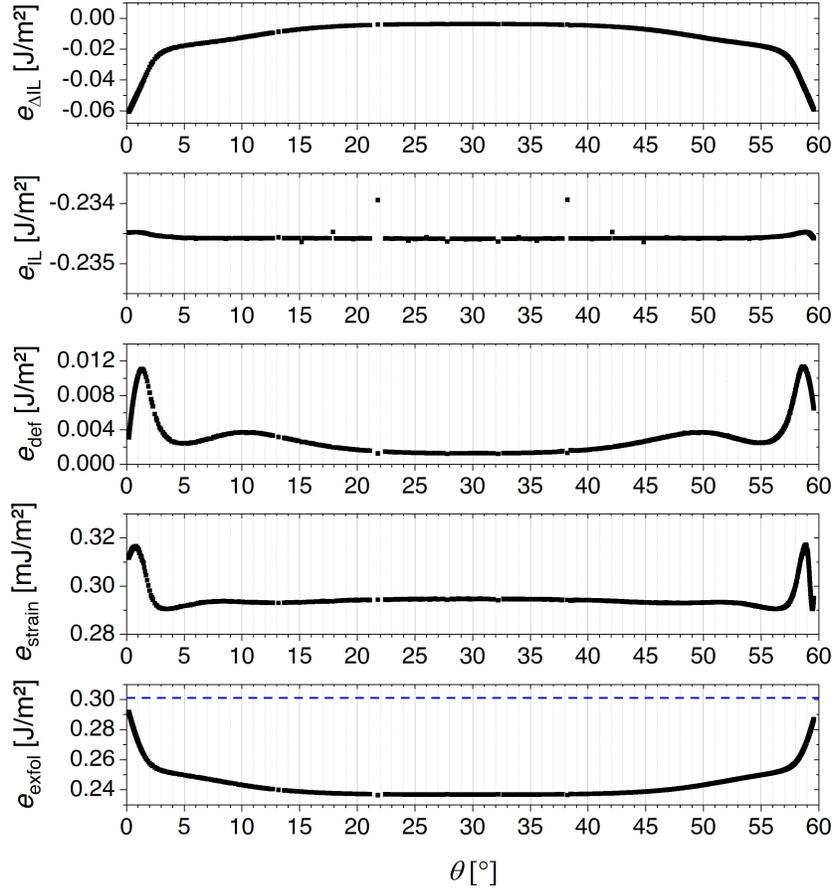

FIG. S16. Exfoliation energy and its energy contributions in tBL $MoS_2$ as a function of $\theta$. Shown are (from top to bottom) the adhesion change due to relaxation $e_{\Delta\mathrm{IL}}$, the adhesion in rigidly twisted systems $E_{\mathrm{IL}}$, the deformation energy $e_{\mathrm{def}}$, the strain energy $e_{\mathrm{strain}}$, and the exfoliation energy $e_{\mathrm{exfol}}$. A dashed blue line indicates the lowest-energy BL $H_h^h$. The values of $e_{\mathrm{exfol}}$ by stacking are $0.301\,\mathrm{J/m^2}$ for $H_h^h$, $0.298\,\mathrm{J/m^2}$ for $R_h^M$, $0.205\,\mathrm{J/m^2}$ for $R_h^h$ and $0.203\,\mathrm{J/m^2}$ for $H_h^M$. The deformation energy has a maximum at $1.4°$ ($58.6°$). Towards $0°$ and $60°$, it decreases towards zero, as the uniform stacking domains dominate the structure.

*Piecewise linear fit of the exfoliation energy*

For $0° < \theta < 30°$, and similarly for $30° < \theta < 60°$, $e_{\mathrm{exfol}}$ shows three distinct linear parts. This was also observed by van Wijk et al. [14] in tBL graphene, who describe a strong decrease of the formation energy for $\theta < 2°$. We performed a piecewise linear fit:

$$e_{\mathrm{exfol}}(\theta) = \begin{cases} e_0 + k_1 \cdot \theta = e_1(\theta) & , \theta < \theta_1 \\ e_1(\theta) + k_2 \cdot (\theta - \theta_1) = e_2(\theta) & , \theta_1 \leq \theta < \theta_2 \\ e_2(\theta) + k_3 \cdot (\theta - \theta_2) & , \theta_2 \leq \theta. \end{cases} \qquad (16)$$

TABLE S5. Piecewise linear fit parameters of $e_{\mathrm{exfol}}$ (see Fig. S17), separated by the intervals $0° < \theta < 30°$ and $30° < \theta < 60°$, including the uncertainties of the curve fit. The parameters $k_i$ and $\theta_i$ are interchanged in the second interval for easier comparison. The value of $\theta_1$ is sensitive to the amount of data points with small $\theta$ available. The value of $\theta_2$, however, is insensitive towards such changes and is, thus, used directly as the transition angle between transition and moiré regime.

| Parameters | $0° < \theta < 30°$ | $30° < \theta < 60°$ |
|---|---|---|
| $e_0$ [mJ/m$^2$] | $295.47 \pm 0.09$ | $231.15 \pm 0.33$ |
| $k_1$ [mJ/m$^2$ per 1°] | $-23.06 \pm 0.16$ | $19.61 \pm 0.15$ |
| $k_2$ [mJ/m$^2$ per 1°] | $-1.42 \pm 0.02$ | $1.40 \pm 0.01$ |
| $k_3$ [mJ/m$^2$ per 1°] | $-0.14 \pm 0.01$ | $0.16 \pm 0.01$ |
| $\theta_1$ [°] | $1.8 \pm 0.01$ | $58.0 \pm 0.01$ |
| $\theta_2$ [°] | $13.4 \pm 0.13$ | $46.8 \pm 0.09$ |
| $R^2$ | $0.999$ | $0.998$ |

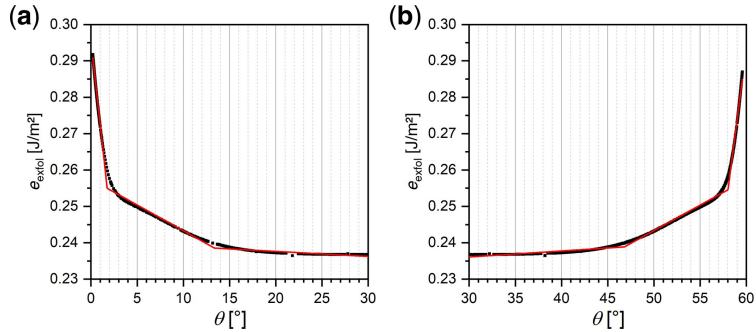

FIG. S17. Piecewise linear fit of the formation energy for (a) $0° < \theta < 30°$ and (b) $30° < \theta < 60°$.

**Displacement field**

The displacement field $\vec{v}(\vec{r})$ [15] is the change of the atomic coordinates $\vec{x}_i$ between the fully relaxed and the initial, rigidly twisted system:

$$\vec{v}(\vec{r}_i) = \vec{x}_{i,\text{relaxed}} - \vec{x}_{i,\text{initial}}. \tag{17}$$

The compression of the cell constant is neglected as it gives only a small relative change in the atomic coordinates Thus, the already lattice-optimized initial structure is used. Furthermore, the out-of-plane (oop) component is neglected as the initial interlayer distance can be chosen arbitrarily, adding ambiguity to its value.

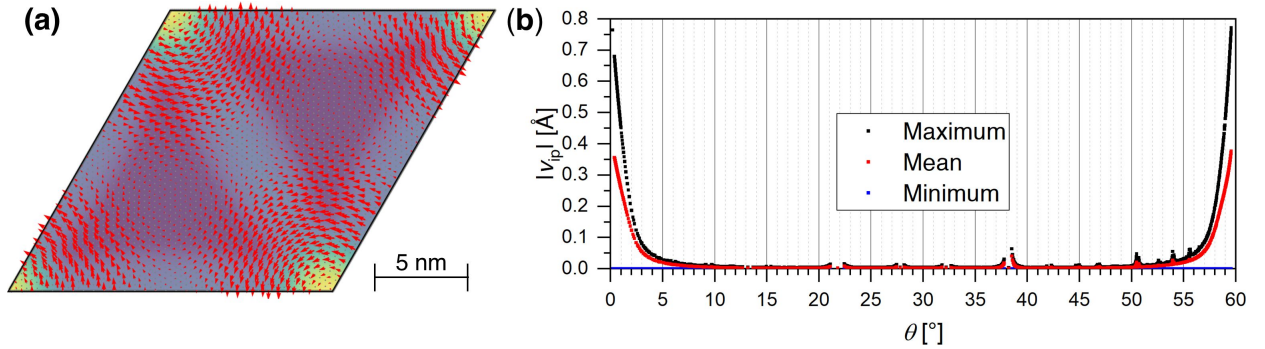

FIG. S18. (a) Displacement field of the Mo atoms inside the moiré cell for an individual layer in tBL MoS$_2$ with $\theta = 1.05°$. The red arrows show the vector field with the magnitude correlating to their length, multiplied by a factor of 20. The interlayer distance is shown in the background. For small $\theta$, the vectors form a circular pattern around the nodes, with the largest amplitude at the boundary between structural elements. They rotate in opposite directions around the nodes in the individual layers to form local high-symmetry stacking areas with an effective local $\theta$ of $0°$. Areas that are already close to a high-symmetry stacking in the rigidly twisted structure can thus be understood as seeds for the atomic reconstruction. (b) Magnitude of the in-plane displacement field $|v_{\text{ip}}|$ of the Mo atoms as a function of $\theta$. It is negligible for $\theta > 3°$ ($\theta < 57°$) when only oop relaxation effects are present. However, it increases strongly near $0°$ and $60°$. This reproduces well the findings already reported in the literature.[14, 16]

**Strain fields**

The isotropic strain field $\epsilon$ is a scalar field:

$$\epsilon_i = \frac{1}{N} \sum_{j=1}^{N} \left| \frac{\vec{d}_{ij}}{d_{\mathrm{ML}}} \right| \qquad (18)$$

with $d_{\mathrm{ML}} = 3.186\,\text{Å}$. The anisotropic strain field $\vec{u}$ is a vector field:

$$\vec{u}_i = \sum_{j=1}^{N} \vec{d}_{ij}. \qquad (19)$$

They use the vector $\vec{d}_{ij}$ from Mo atom $i$ to its next neighbor, $j$, and $N = 6$ as the number of next neighbors. The isotropic strain field can be understood as the local variation of $a_{\mathrm{prim}}$ relative to the unstrained ML. Inside the moiré regime, it shows a situation similar to a sinusoidal modulation as described by van Wijk et al.[14]. The anisotropic strain field $\vec{u}$ shows the local variation and the direction in which the Mo-Mo distances change the strongest.

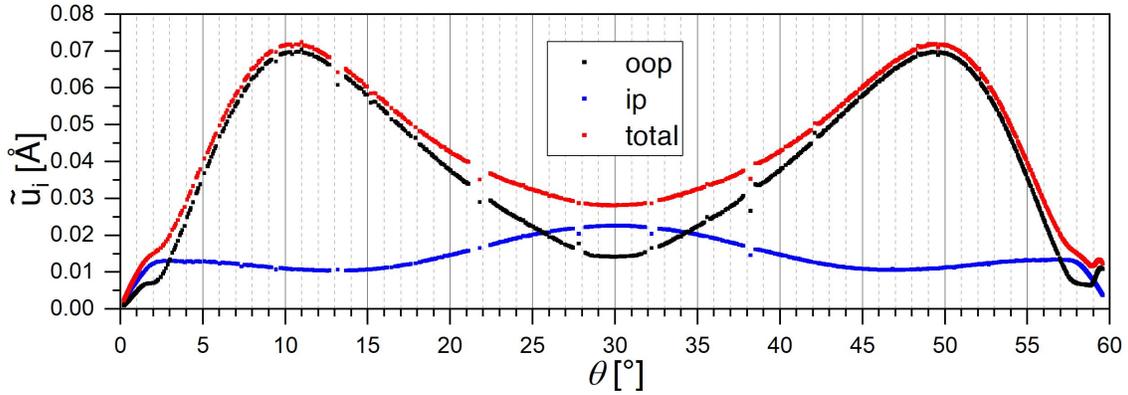

FIG. S19. Anisotropic strain field $\tilde{u}$, averaged over all atoms inside a layer, as a function of $\theta$. The ip, oop components, and the absolute values of the full anisotropic strain field vectors are shown. For most $\theta$, the oop component dominates as the main deformation of the layers is in the $z$-direction. This only changes for $\theta < 3°$ ($\theta > 57°$), where the ip component dominates due to the increasing ip relaxation and the formation of extended stacking domains with only minor oop anisotropy. The order of the anisotropic strain field vanishes with increasing $\theta$ around 12° (48°).

**Comparison between $\theta \to 0°$ and $\theta \to 60°$**

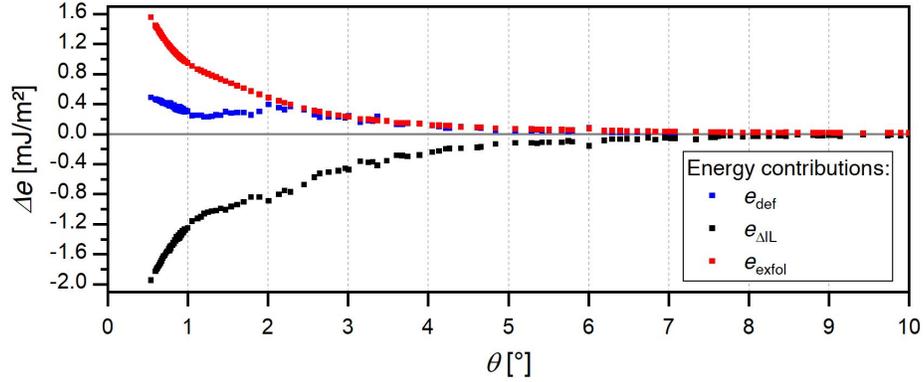

FIG. S20. Difference of the energy contributions between $\theta \to 0°$ and $\theta \to 60°$, calculated by $\Delta e(\theta) = e(60° - \theta) - e(\theta)$. The values are generally small, with a maximum deviation for $e_{\Delta IL}$ of $\approx 1.9 \, \mathrm{mJ/m^2}$ ($\approx 0.6\%$). Structures with $\theta \to 60°$ are, thus, stronger bound. The reason is the more extended stacking domains that increase the adhesion and overcompensate the energy needed to deform the solitons. Values for $\theta > 10°$ are not shown as the deviations become negligible. Both $e_{strain}$ and $e_{IL}$ are excluded as their $\Delta e$ values are very small ($e_i < 0.01 \, \mathrm{mJ/m^2}$).

**Influence of inherent interlayer strain**

A few exemplary structures with up to 2% strain were analyzed to gain qualitative insight into the effect of inherent interlayer strain. It originates from the mismatch in the cell constant of the individual layers, similar to the situation on a substrate or in vdW heterobilayers. We observe only small structural changes. The most prominent is the formation of bent solitons near 0°. The solitons form a vortex-like pattern around the nodes (both clockwise and counterclockwise, but always identical for all solitons inside one structure).

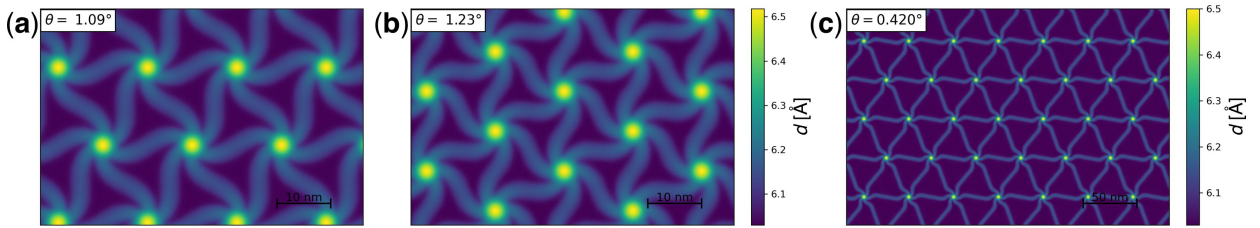

FIG. S21. Local interlayer distance $d$ in tBL $MoS_2$ structures with inherent interlayer strain. Subplots (a) and (b) share the same color bar.

**TRANSITION BETWEEN STRUCTURAL REGIMES**

Changes in the properties of the tBL MoS$_2$ structures accumulate at specific values of $\theta$. These are used as the boundary between the structural regimes. The transition angle $\theta = 13 \pm 1°$ ($\theta = 47 \pm 1°$) is chosen using

- the parameter $\theta_2$ of the fit of $e_{\text{exfol}}$,

- the vanishing stacking domains in the geometric model for $14° - 48°$,

- the isotropic strain field losing its order for $12° - 48°$, and

- the nodes vanish in the local interlayer distance plots for $12° - 46°$.

The transition angle $\theta = 6°$ ($\theta = 54°$) is chosen using the value of $\theta$ at which

- $d_{\text{max}}$ reaches the value of the S@S stacking type and

- the displacement field around the nodes emerges.

The transition angle $\theta = 3°$ ($\theta = 57°$) is chosen because here

- the height of nodes and solitons reaches its maximum,

- the moiré periodicity $a_{\text{M}}$ and base width of nodes decouple,

- $d_{\text{min}}$ reaches the value of the S@Mo stacking type,

- $d_{\text{mean}}$ starts to strongly decrease, and

- the ip component becomes the dominating component of the anisotropic strain field.

## ELECTRONIC PROPERTIES

### Band features at small twist angles

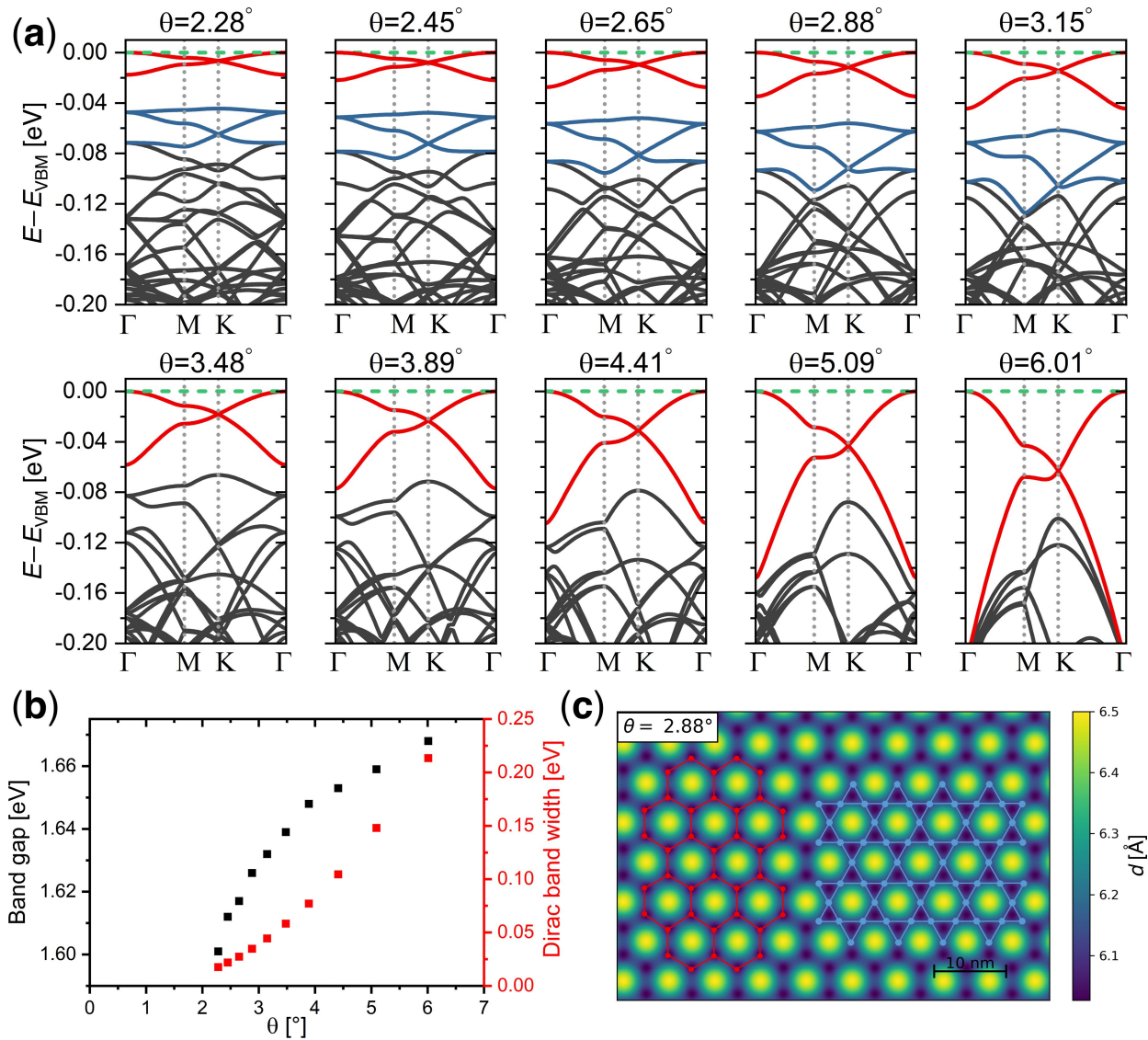

FIG. S22. (a) Band structure for small $\theta$, calculated at DFTB level, showing the occupied bands. Red lines highlight the honeycomb bands with a Dirac cone at $K$, while blue highlights kagome-type bands [17] emerging for $\theta < 3°$. (b) Value of the indirect band gap and the total bandwidth of the honeycomb bands (highlighted in red) as a function of $\theta$. (c) Visualization of the stacking domains forming a honeycomb pattern (red), while the solitons for a kagome lattice (blue).

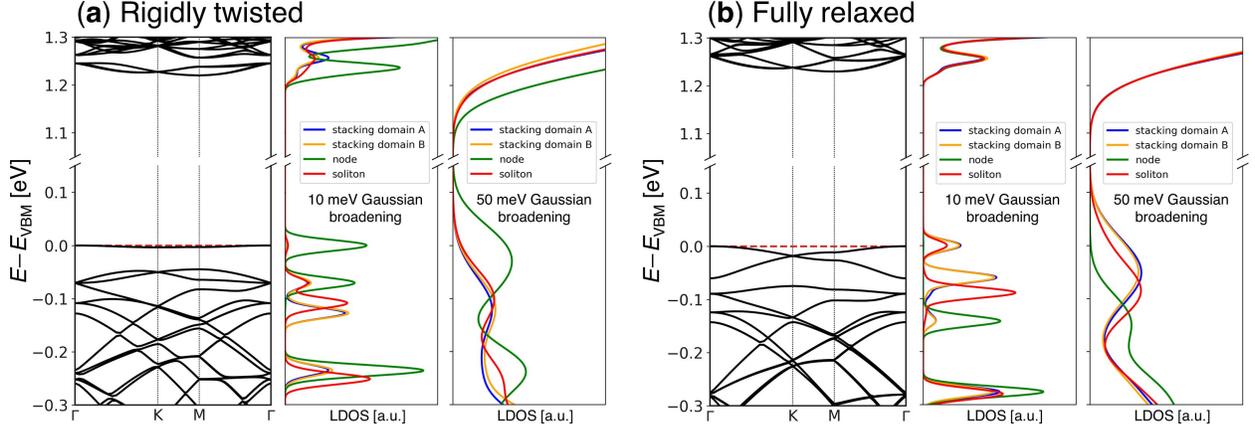

FIG. S23. Atom-projected DOS (LDOS) of the Mo atoms inside an individual layer of tBL MoS$_2$ with $\theta = 3.89°$, calculated at the DFT level using (a) a rigidly twisted and (b) a fully relaxed structure. The LDOS is shown for Mo atoms in the center of the structural elements. In the fully relaxed structure, contributions from the stacking domains dominate the honeycomb bands. Contributions from the solitons dominate the bands below, which become the kagome-type bands at smaller $\theta$. This links the emergence of kagome-type bands to a kagome lattice formed by the center of the solitons (see Fig. S22(c)).

**Band gap vs. interlayer distance**

The band gap $E_{\text{gap}}$ shows a linear dependence on the mean interlayer distance $d_{\text{mean}}$:

$$E_{\text{gap},\theta<30°}[\text{eV}] = 1.7641 \cdot d_{\text{mean}}[\text{Å}] - 9.8008, R^2 = 0.991,$$

$$E_{\text{gap},\theta>30°}[\text{eV}] = 1.5595 \cdot d_{\text{mean}}[\text{Å}] - 8.4978, R^2 = 0.991.$$

However, no structures from the domain-soliton regime are included, and the functions do not reproduce the values of the MoS$_2$ BL stackings, so this linear dependence cannot be applied universally.

**Local gap**

TABLE S6. Energy of the CBM, VBM, and local gap in tBL MoS$_2$ with $\theta = 3.89°$ in the case of (left column) rigidly twisted layers and (right column) a fully relaxed structure. The atomistic structure is shown as an overlay in the local gap plots. Areas of high (low) VBM energy correspond to areas with small (large) local band gap. The variation of the CBM is small compared to the VBM and can be treated as constant in the fully relaxed structure (changes below 2 meV). Using the atom-projected DOS of the S atoms located on the inside of the tBL system instead of the Mo atoms would result in qualitatively identical results.

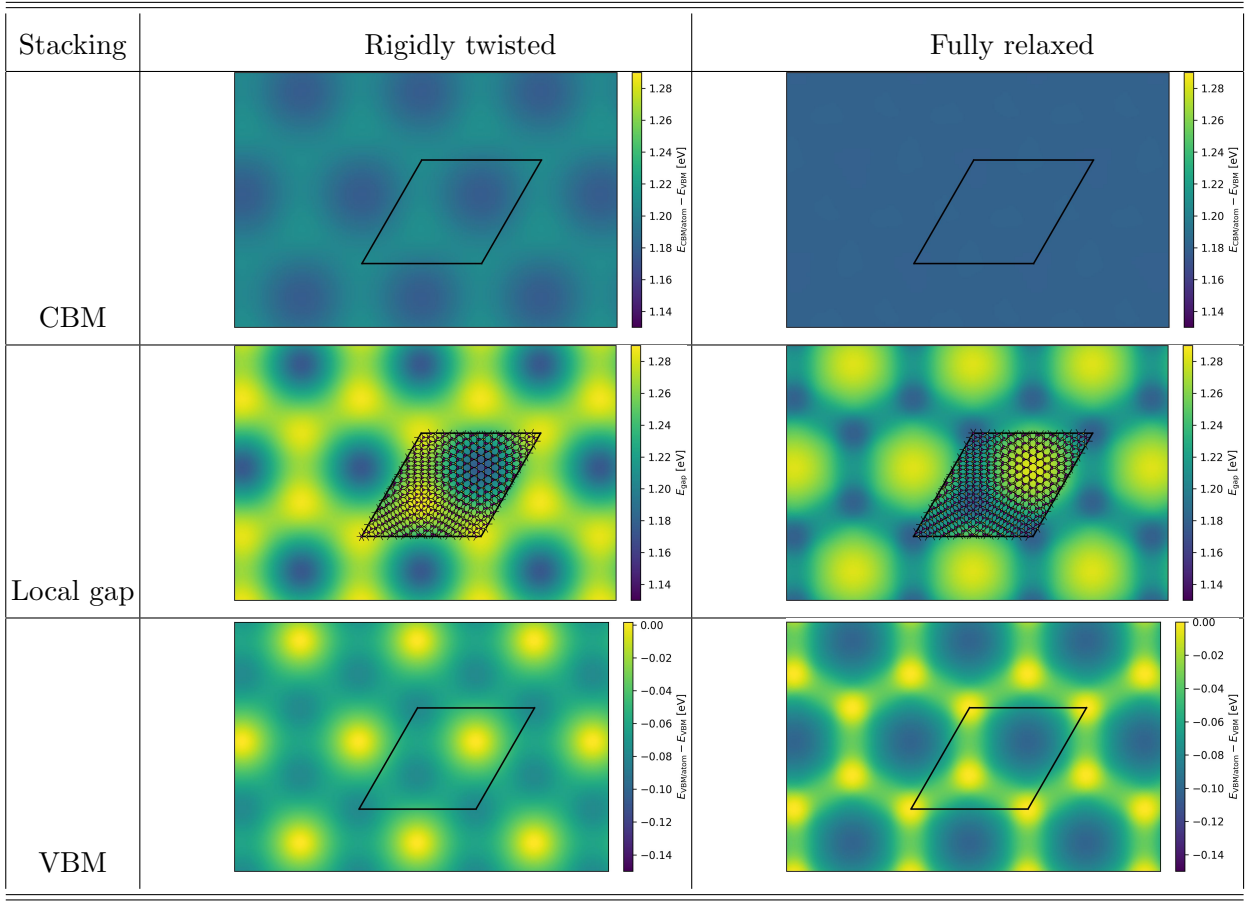

TABLE S7. Energy of the CBM, VBM, and local gap in tBL MoS$_2$ with $\theta = 56.11°$ in the case of (left column) rigidly twisted layers and (right column) a fully relaxed structure. The atomistic structure is shown as an overlay in the local gap plots. The behavior for $\theta \to 60°$ agrees qualitatively with $\theta \to 0°$. The only difference is that adjacent stacking domains alternate between low and intermediate values of local gap and VBM energy, due to different local stacking inside the stacking domains.

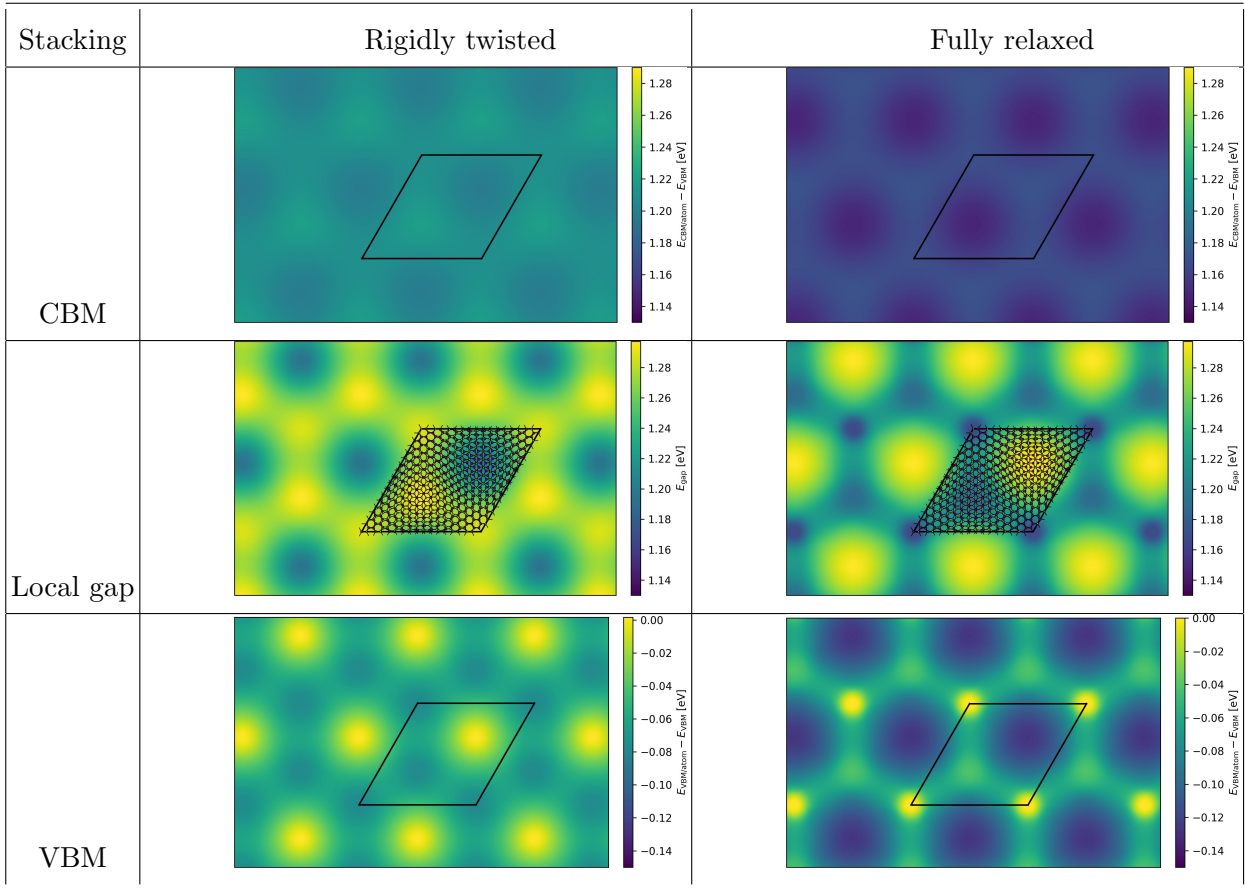

## LDOS at the VBM

TABLE S8: LDOS at the energy of the VBM in rigidly twisted and fully relaxed tBL MoS$_2$ structures for different $\theta$. The atomic structure inside the moiré cell is shown. The color bar is identical for all subplots. The qualitative patterns do not change with $\theta$. However, the modulation of the LDOS decreases for $\theta \to 30°$ and vanishes inside the moiré regime. Close to 60°, the different local stacking inside adjacent stacking domains are again visible with different LDOS values.

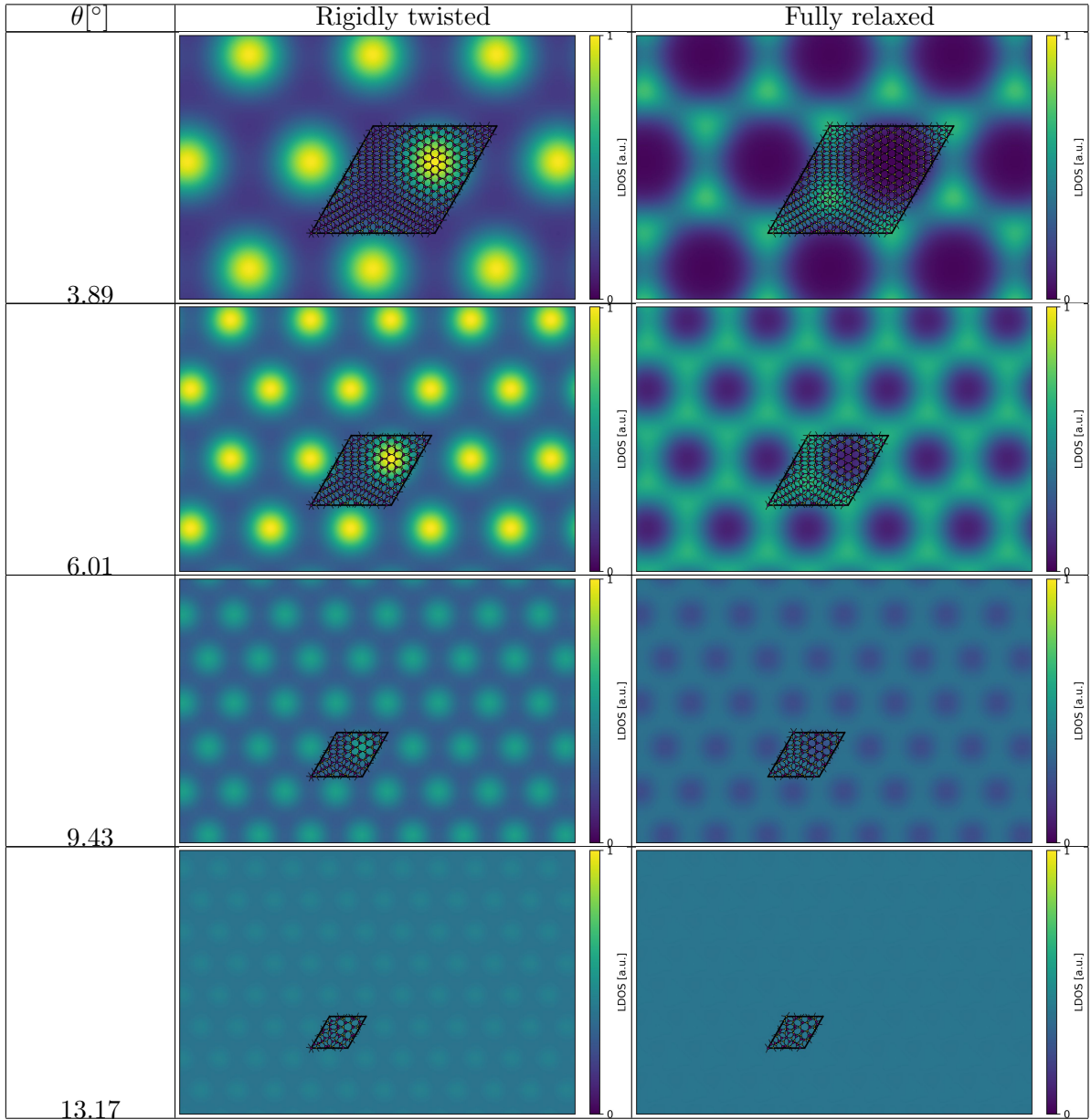

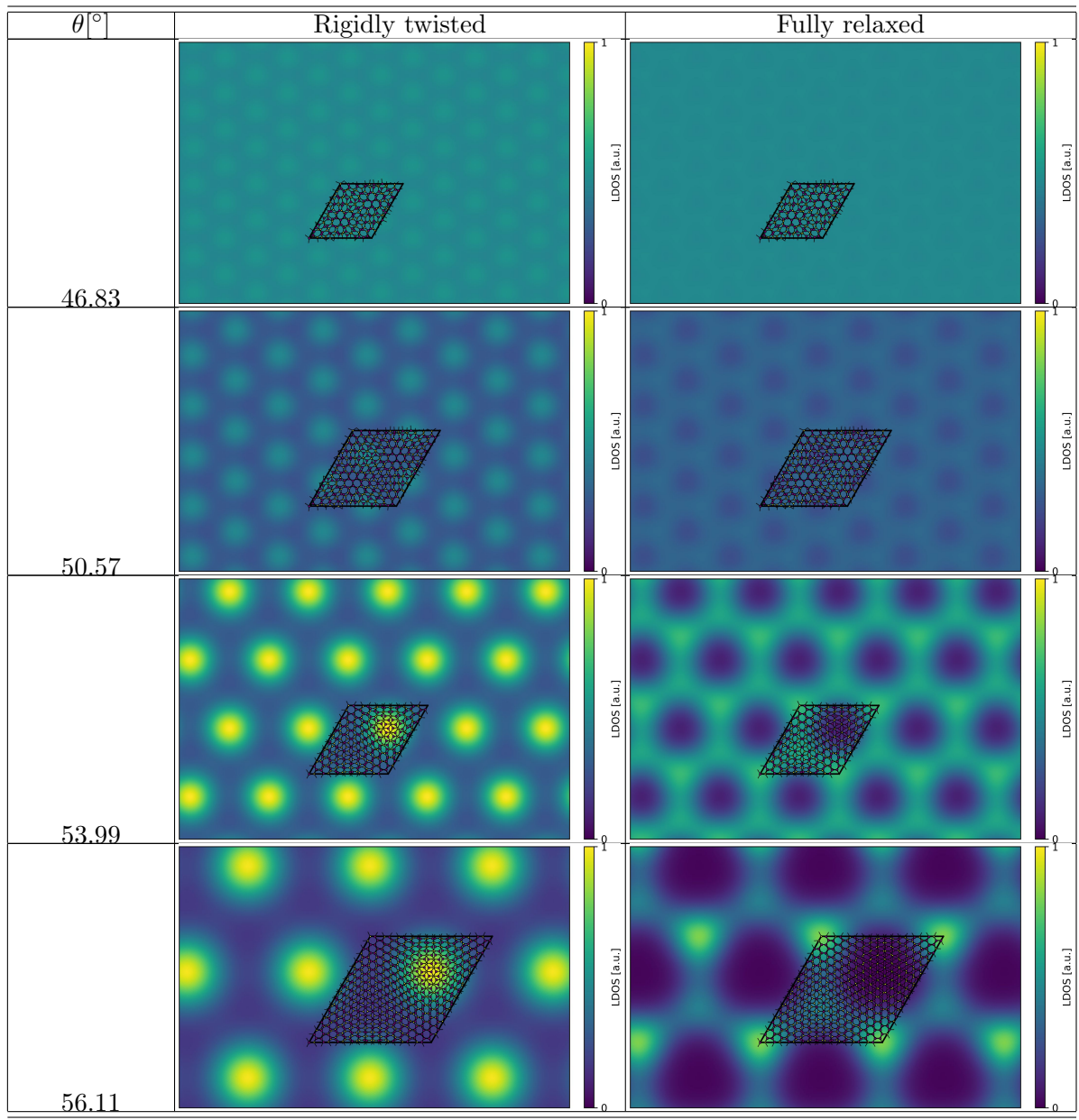




\begin{thebibliography}{62}
\bibitem [{ZEN()}]{ZENODO}%
  \BibitemOpen
  \href@noop {} {}\bibinfo {note} {The calculation files, the rigidly twisted
  and fully ReaxFF-optimized structures, and movies showing the change of the
  interlayer distance landscape and the strain fields with the twist angle are
  available at ZENODO. URL: doi.org/10.5281/zenodo.7243735}\BibitemShut
  {NoStop}%
\end{thebibliography}
\end{document}